\documentclass[namedreferences]{solarphysics}
\usepackage[hyperref,optionalrh,solaromanenum]{spr-sola-addons} 
\usepackage{graphicx}                    
\usepackage{color}                       
\usepackage{breakurl}                    
\usepackage{pdflscape}					 

\chardef\us=`\_

\begin{document}

\begin{article}

\begin{opening}

\title{Connecting Coronal Mass Ejections to their Solar Active Region Sources: Combining Results from the HELCATS and FLARECAST Projects}

%
\author[addressref={aff1},corref,email={sophie.murray@tcd.ie}]{\inits{S. A. }\fnm{Sophie A. }\lnm{Murray}\orcid{0000-0002-9378-5315}}
\author[addressref={aff2},corref]{\inits{J. A. }\fnm{Jordan A.}~\lnm{Guerra}\orcid{0000-0001-8819-9648}}
\author[addressref={aff3},corref]{\inits{P. }\fnm{Pietro}~\lnm{Zucca}}
\author[addressref={aff4},corref]{\inits{S.-H. }\fnm{Sung-Hong}~\lnm{Park}\orcid{0000-0001-9149-6547}}
\author[addressref={aff1,aff5},corref]{\inits{E. P. }\fnm{Eoin P.}~\lnm{Carley}}
\author[addressref={aff1},corref]{\inits{P. T. }\fnm{Peter T.}~\lnm{Gallagher}\orcid{0000-0001-9745-0400}}
\author[addressref={aff5,aff6},corref]{\inits{N. V. }\fnm{Nicole}~\lnm{Vilmer}}
\author[addressref={aff7},corref]{\inits{V. }\fnm{Volker}~\lnm{Bothmer}}

%
\runningauthor{S.A. Murray \textit{et al.}}
\runningtitle{Combining results from HELCATS and FLARECAST}

\address[id={aff1}]{Astrophysics Research Group, School of Physics, Trinity College Dublin, Ireland}
\address[id={aff2}]{Department of Physics, Villanova University, Villanova PA, USA}
\address[id={aff3}]{ASTRON Netherlands Institute for Radio Astronomy, Postbus 2, 7990 AA Dwingeloo, The Netherlands}
\address[id={aff4}]{Institute for Space-Earth Environmental Research (ISEE), Nagoya University, Japan}
\address[id={aff5}]{LESIA, Observatoire de Paris, PSL Research University, CNRS, Sorbonne Universit\`{e}s, UPMC Univ. Paris 06, Univ. Paris Diderot, Sorbonne Paris Cit\`{e}, 5 place Jules Janssen, 92195 Meudon, France}
\address[id={aff6}]{Station de Radioastronomie de Nancay, Observatoire de Paris, PSL Research University, CNRS, Univ. Orle\`{a}ns, 18330 Nancay, France}
\address[id={aff7}]{Institute of Astrophysics, University of G\"ottingen, Germany}

\begin{abstract}
Coronal mass ejections (CMEs) and other solar eruptive phenomena can be physically linked by combining data from a multitude of ground-based and space-based instruments alongside models, however this can be challenging for automated operational systems. The EU Framework Package 7 HELCATS project provides catalogues of CME observations and properties from the \textit{Heliospheric Imagers} onboard the two NASA/STEREO spacecraft in order to track the evolution of CMEs in the inner heliosphere. From the main HICAT catalogue of over 2,000 CME detections, an automated algorithm has been developed to connect the CMEs observed by STEREO to any corresponding solar flares and active region (AR) sources on the solar surface. CME kinematic properties, such as speed and angular width, are compared with AR magnetic field properties, such as magnetic flux, area, and neutral line characteristics. The resulting LOWCAT catalogue is also compared to the extensive AR property database created by the EU Horizon 2020 FLARECAST project, which provides more complex magnetic field parameters derived from vector magnetograms. Initial statistical analysis has been undertaken on the new data to provide insight into the link between flare and CME events, and characteristics of eruptive ARs. Warning thresholds determined from analysis of the evolution of these parameters is shown to be a useful output for operational space weather purposes. Parameters of particular interest for further analysis include total unsigned flux, vertical current, and current helicity. The automated method developed to create the LOWCAT catalogue may also be useful for future efforts to develop operational CME forecasting.
\end{abstract}

%

\end{opening}

%
\section{Introduction}\label{s:intro} 

Severe space weather has the potential to significantly impact a range of vital technologies, both on Earth and in near-Earth space. Adverse space weather is known to result from solar eruptions in the form of solar flares ,and coronal mass ejections (CMEs), as well as associated solar energetic particle events. These eruptions often originate from turbulent and highly complex active region (AR) magnetic fields. Understanding the processes involved in these solar eruptions, as well as their sources, is imperative to provide accurate space weather forecasts. CMEs are a main focus of current space weather operational systems considering their potentially severe impact on Earth, such as communication and navigation system disruption, power blackouts, and damage to spacecraft instrumentation \citep[see \textit{e.g.}][for an economic perspective]{eastwood17}. There are considerable research efforts underway to improve our understanding of CMEs to feedback to operational space weather forecasting efforts.

Catalogues of CME events have proven useful in order to better study their propagation. The NASA CDAW Data Center catalogue \citep{gopalswamy09} contains manually identified CMEs since 1996 from the \textit{Large Angle and Spectrometric Coronagraph} (LASCO) onboard the \textit{Solar and Heliospheric Observatory} \citep[SOHO;][]{brueckner95} spacecraft. Automated catalogues such as CACTUS \citep{robbrecht04} and CORIMP \citep{byrne12,morgan12} have become increasingly popular in recent years, considering the difficult task of manually identifying the significant number of CME events that are observed in an age of improved availability of spacecraft observations. Much previous work has been undertaken to statistically analyze properties of the CMEs made available by these catalogues \citep[see][and references therein]{webb12}.

Most CME catalogues currently available have made use of the LASCO coronagraphs. However since 2007 the \textit{Solar TErrestrial RElations Observatory} \citep[STEREO;][]{kaiser08} mission has provided an unprecedented insight into CME propagation, including a 3D perspective when the two STEREO A and B spacecraft were in the correct orbital position. Wide-angle imaging of the inner heliosphere has revolutionised our understanding of CMEs, solar wind, and co-rotating interaction regions \citep{harrison17}. Heliospheric imaging provides the ability to track the evolution of these phenomena as they propagate through the inner heliosphere. Heliospheric Cataloguing, Analysis and Techniques Service (HELCATS) is a European Union Framework Package 7 project that provides the definitive catalogue of CME events observed by the STEREO \textit{Heliospheric Imagers} \citep[HI;][]{eyles09}.

Thus far the results of the HELCATS project have been successful in enhancing our understanding of CMEs and their propagation \citep[see \textit{e.g.}][]{plotnikov16, mostl17, rouillard17}. The results are particularly useful for improving CME arrival time predictions at Earth, and even at other planets in our solar system. However, forecasting when the CME itself will erupt remains elusive \citep{zheng13}. The more established field of flare forecasting now uses an extensive variety of methods to create predictions, from simple Poisson statistics \citep{bloomfield12} to more complex machine learning methods \citep{ahmed13}. Unfortunately it is still unclear what methods are the best or even better than climatology \citep{barnes16}, and many operational forecasting centers still rely on human forecasters to tailor predictions \citep{murray17}. Flare Likelihood And Region Eruption foreCASTing (FLARECAST) is a European Union Horizon 2020 project that aims to solve this uncertainty by creating an automated system that forecasts flares with unprecedented accuracy. In order to achieve such a goal, FLARECAST has created a large database of flaring activity predictors related to their AR sources and is undertaking a comprehensive evaluation of the many advanced prediction algorithms.

While there are a considerable number of different flare forecasting methods available, the vast majority of them use very similar predictors, namely some measure of the ``complexity'' of the solar surface AR magnetic field. It is natural to expect that AR magnetic fields may also prove to be key for CME forecasting purposes. Previous work has investigated correlations between AR photospheric magnetic field and CME properties \citep[\textit{e.g.}][]{ventakrishnan03, guo07, park12, lee15, tiwari15}, as well as studied the locations of CME sources on the solar disk \citep{lara08, yashiro08, wang11}. The connection between CMEs and associated flares has also been extensively studied, particularly between the duration and intensity of the flares and CME speed \citep[\textit{e.g.}][]{harrison95, andrews03, youssef12, harra16}. Statistical analyses of these AR properties are important to investigate the solar surface source of CMEs in order to make the first steps towards CME forecasting, just as the field of flare forecasting has done before it. However this work is limited by the observations available at the time of study and the considerable time it takes to manually associate the CMEs in these large databases with flares and ARs.

The main deliverables of the HELCATS project are focused on creating catalogue parameters related to STEREO CME observations in the heliosphere, however the aim of HELCATS Deliverable 3.4 is to develop an automated algorithm to determine the solar surface source of the CME events. The algorithm is used to create the HELCATS LOWCAT catalogue, linking the STEREO/HI-observed CMEs to observations of associated solar flares and ARs, if any. This HELCATS catalogue provides an excellent opportunity for cross-collaboration with the FLARECAST project and their database of AR properties. It is worth noting that these projects focus on associations between CMEs and flares (and their AR sources). However CMEs can be also associated with other phenomena such as filament or prominence eruptions, with or without flares \citep[see \textit{e.g.}][]{moon02, jing04}.

This paper outlines the main algorithm steps in creating the LOWCAT catalogue, and the first initial scientific investigations using CME, flare, and AR data from both HELCATS and FLARECAST projects are also described. Section~\ref{s:lowcat} describes the algorithm and resulting catalogue properties in detail, and highlights some initial results. Section~\ref{s:flarecast} outlines the FLARECAST AR property database, and presents some example property comparisons between the two project databases. Discussion and conclusions are found in Section~\ref{s:concl}, particularly highlighting the availability of the data to the community and encouraging its use for future research efforts.


\section{The LOWCAT Catalogue}\label{s:lowcat} 

\subsection{Algorithm Development}
With over 2,000 CME observations from mid 2007 until early 2017 available for analysis in the HICAT catalogue developed in HELCATS Work Package 2\footnote{see \url{https://www.helcats-fp7.eu/catalogues/wp2\_cat.html}.} \citep{barnes15}, manually identifying each solar surface source for that many events was not feasible. To that end, an automatic algorithm has been developed that uses back-propagation to identify flare events and AR sources correlated with the CME events listed. The algorithm first associates the HI events to COR2 observations closer to the Sun, then to solar flare events, and finally to ARs on the solar surface. Note that while this simple method is sufficient for the purposes of this work, more complex methods could be used in future developments \citep[see \textit{e.g.}][]{davies13}. Thomson scattering effects and 3D geometries need to be taken into account for events observed far away from the solar surface \citep{vourlidas06, davies12}. This is particularly important for observations by HI-2, however only HI-1 observations from either STEREO A or B were used to create the HICAT catalogue.

Assuming constant radial velocity and constant CME width, a simple ballistic propagation model is used to search for candidates. First the algorithm identifies a time window during which a STEREO/HI-observed CME might have been observed by the STEREO/COR2 coronagraph. Here the initial distance can be taken as 12~R$_{\odot}$, final distance 2~R$_{\odot}$, and a typical range of CME speeds (150-1500~km) defined by \citet{yurchyshyn05} is used to constrain the search. The CACTUS database, which contains CME events identified by an automated method, is searched for events occurring during this time window. The database was originally developed for LASCO/C2 observations, however was extended to include STEREO data during the HELCATS project \citep{pant16}. The algorithm selects a COR2-observed CME with the closest angular width to the HI-observed CME, constrained by the north and south position angles.

Ballistic propagation is again used to define a time window of possible associated flares, using the identified COR2 CME event speeds where available. Here, the initial distance is taken as 2~R$_{\odot}$ and final distance can be set as high as 0.5~R$_{\odot}$ to take into account the non-constant speed of the initial CME phase \citep{byrne10}. The algorithm searches for flare events in the National Oceanic and Atmospheric Administration Space Weather Prediction Center (NOAA/SWPC) Edited Solar Event Lists\footnote{see \url{ftp://ftp.swpc.noaa.gov/pub/warehouse}.} by default. It will also check the NASA \textit{Reuven Ramaty High Energy Solar Spectroscopic Imager} (RHESSI) flare list\footnote{see \url{http://hesperia.gsfc.nasa.gov/hessidata/dbase/hessi\_flare\_list.txt}.} if no events are found in the NOAA flare list, and for those events without any given location. The resulting flare list is further constrained by solar hemispheric location, ensuring the CME position angle corresponds to the quadrant where the potential flare peaks are located. If multiple flares are found within the search window the closest flare to the time window start is selected. The algorithm also limits the output to flares of \textit{Geostationary Operational Environmental Satellite} (GOES) magnitude B-class and above, and for the RHESSI event list it allows only confirmed flare detections with a high-quality level defined in the list flags\footnote{see \url{http://sprg.ssl.berkeley.edu/$\sim$jimm/hessi/hsi\_flare\_list.html}.}. 

Finally, the algorithm associates any identified flare events with corresponding NOAA-numbered ARs. The flare peak location is used to search for nearby ARs on the solar disk if no number has been listed in the flare database. Some basic properties of these AR sources can be obtained from the NOAA/SWPC Solar Region Summaries, including the Hale \citep[particularly the Modified Mount Wilson which includes $\delta$ spots;][]{kunzel65} and McIntosh \citep{mcintosh90} classifications.


\subsection{Solar Monitor Active Region Tracker}
\begin{figure}[!t]
\centering
\noindent\includegraphics[width=\textwidth]{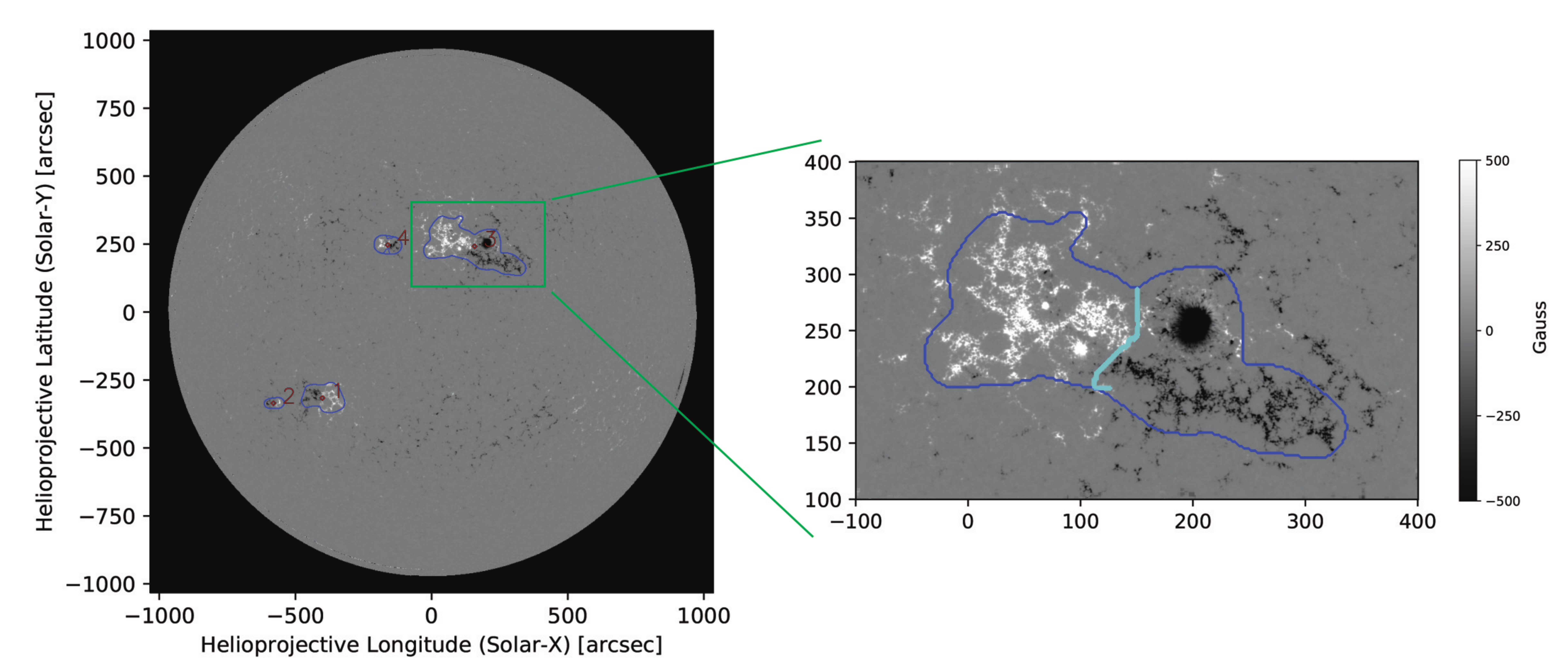}
\caption[SMART detections for a HMI line-of-sight magnetogram on 21 June 2011.]{SMART detections for a HMI line-of-sight magnetogram on 21 June 2011 04:00 UTC (left). The zoomed-in region (right; green box on left) shows detection number 3, which corresponds to NOAA AR number 11236, and was the source of a C-class flare on 21 June 2011 as listed in the LOWCAT database. The dark blue line indicates the detection boundary, and light blue line the PIL.}
\label{fig:smart}
\end{figure}

In order to complete the HELCATS database, the Solar Monitor Active Region Tracker algorithm \citep[SMART;][]{higgins11} is then run to obtain magnetic field properties of the identified ARs. SMART is an automated system for detecting, tracking, and cataloguing ARs throughout their evolution. The algorithm uses consecutive image differencing to remove quiet-Sun and transient magnetic features, then region-growing techniques to group flux concentrations into classifiable features. See \citet{higgins11} for more details about how the algorithm works. Figure~\ref{fig:smart} shows an example of a SMART detection on 21 June 2011. The dark blue lines in the Figure indicate the boundaries of the SMART AR detections, within which various magnetic field properties are calculated.

SMART is run on regions using SOHO's \textit{Michelson Doppler Imager} \citep[MDI; ][]{scherrer95} and \textit{Solar Dynamics Observatory Helioseismic and Magnetic Imager} \citep[SDO/HMI; ][]{scherrer12} line-of-sight (LOS) magnetograms, depending on the date of the observations (HMI data being available from mid 2010). The flare peak time and location can be used to check for SMART regions even if no NOAA regions were identified. SMART calculates more complex magnetic properties related to the polarity inversion line (PIL), including PIL length, bipole separation, \textit{R} value \citep{schrijver07} and WL\textsubscript{sg} \citep{falconer08}. Note that the gradient-weighted integral length of neutral line, WL\textsubscript{sg}, is defined as the line integral of the vertical field horizontal gradient over all PIL intervals where the value of potential horizontal field is greater than 150 gauss. This threshold is similarly used for the \textit{R} value, which describes total unsigned flux near a strong-field, high-gradient PIL. A full list of AR information outputted by the algorithm is shown in Table~\ref{tbl:smart_properties}.

\begin{landscape}
\begin{table}
\caption{Magnetic source active region properties obtained from the SWPC Solar Region Summary and SMART algorithm outputs.}\label{tbl:smart_properties}
\begin{tabular}{l l l l}     
\hline
Source	& Property		& Unit						& Description												\\
\hline
NOAA	& srs\_no			& -								& SWPC-determined active region number of candidate region		\\
NOAA	& srs\_mcintosh	& -								& SWPC-determined McIntosh classification of the group			\\
NOAA	& sr\_hale			& -								& SWPC-determined Hale classification 						\\
NOAA	& srs\_area		& millionths of a solar hemisphere		& SWPC-determined active region total corrected area			\\
NOAA	& srs\_ll			& heliographic degrees				& SWPC-determined longitudinal extent						\\
NOAA	& srs\_nn			& -								& SWPC-determined total number of visible sunspots in the group	\\
SMART	& smart\_totflx		& maxwell							& Total magnetic flux of SMART region						\\
SMART	& smart\_posflx		& maxwell							& Total magnetic flux in positive polarity part of SMART region		\\
SMART	& smart\_negflx		& maxwell							& Total magnetic flux in negative polarity part of SMART region		\\
SMART	& smart\_frcflx		& -								& Flux fraction of SMART region, \textit{i.e.} (posflx - negflx) / totflx	\\
SMART	& smart\_totarea	& millionths of a solar hemisphere		& Total magnetic area of SMART region						\\
SMART	& smart\_posarea	& millionths of a solar hemisphere		& Total magnetic area of positive polarity part of SMART region		\\
SMART	& smart\_negarea	& millionths of a solar hemisphere		& Total magnetic area of negative polarity part of SMART region	\\
SMART	& smart\_bmin		& gauss							& Total negative magnetic field strength of SMART region			\\
SMART	& smart\_bmax		& gauss							& Total positive magnetic field strength of SMART region 			\\
SMART	& smart\_bmean	& gauss							& Mean magnetic field strength of SMART region				\\
SMART	& smart\_psllen		& megameters						& Polarity inversion line length of SMART region				\\
SMART	& smart\_bipolesep	& megameters						& Bipolar separation of SMART region						\\
SMART	& smart\_rvalue		& maxwell							& R value of SMART region								\\
SMART	& smart\_wlsg		& gauss							& WL\textsubscript{sg} value of SMART region					\\
\hline
\end{tabular}
\end{table}
\end{landscape}


\subsection{Resulting Catalogue}
\begin{figure}[!t]
\centering
\noindent\includegraphics[width=\textwidth]{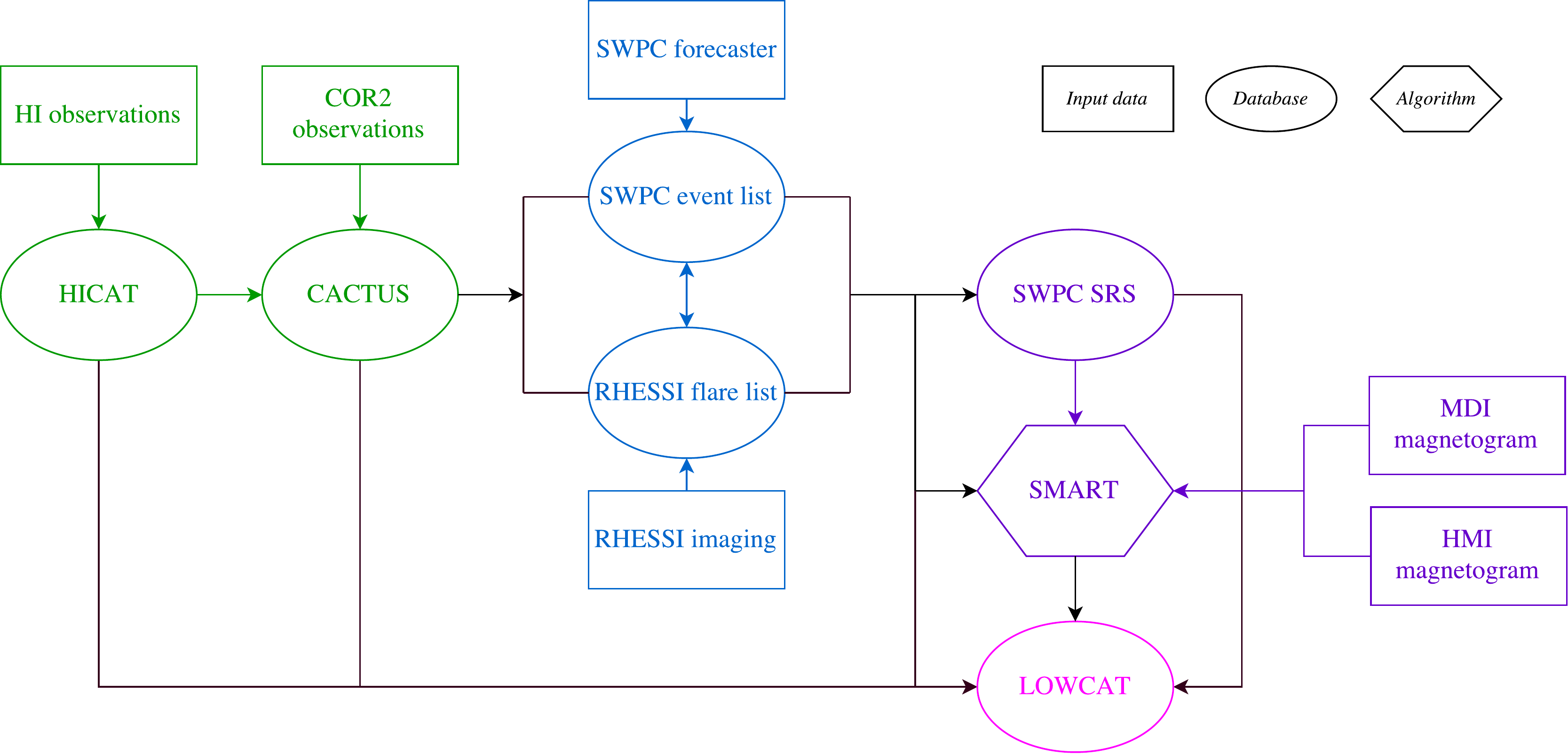}
\caption[LOWCAT flowchart]{Flowchart outlining the steps involved in the automated algorithm used to create the LOWCAT catalogue. The event lists used are identified by ovals, input observations by rectangles, and the SMART algorithm by a diamond. Colours indicate steps in the main process, namely green for CME association, blue for flare association, purple for active region association, and magenta indicates the resulting LOWCAT catalogue.}
\label{fig:flowchart}
\end{figure}

All steps involved in the automated algorithm used to create the catalogue of properties are summarised by the flowchart in Figure~\ref{fig:flowchart}. The resulting LOWCAT catalogue is freely available online at figshare.com\footnote{see https://figshare.com/articles/HELCATS\_LOWCAT/4970222 .} \citep{murray17a}, which is also linked from the HELCATS website. The algorithm has been run with version 4 of the HICAT catalogue, which contains 2,020 HI-observed CMEs, the first of which launched on 15 April 2007 and the last on 26 February 2017. In total there are 1,591 COR2 events, 714 flare events, 552 NOAA regions, and 451 SMART regions in the LOWCAT catalogue besides the original 2,020 HI events. It is worth noting the SMART algorithm has also been run for the Framework Package 7 AFFECTS project KINCAT catalogue\footnote{see https://www.helcats-fp7.eu/catalogues/wp3\_kincat.html .} as part of HELCATS Deliverable 3.3, which contains a small selection of manually identified CME events. The same properties listed in Table~\ref{tbl:smart_properties} are provided as well as an estimation of the tilt of the PIL.

The entries in the LOWCAT catalogue for every 50th event in the HICAT catalogue were manually inspected for verification purposes, checking that the correct COR2 observation (85\% correct), flare event (87.5\% correct), and AR (95\% correct), if any, were listed. Missed CME and flare entries were generally due to being outside the time search window. Limb events were the only ones missed by the AR detection part of the algorithm for the events examined, although these could not have been analysed even if they were included. The results of this basic validation are positive for the automated method, and highlights the feasibility of such a simple method for big data analysis. The CME detection accuracy is comparable to the validation efforts for the CACTUS database, which reproduced $\approx$~75\% of the catalogued CMEs \citep{berghmans02}, although more events were analysed in this study.

The first part of the LOWCAT catalogue contains the ID of the HELCATS HICAT event, for example, the HCME\_B\_\kern-.14ex\_20120305\_01 event, which has been previously been studied by \citet{magdalenic14}, as well as the position and launch time of CME. The automated algorithm identified a COR2 CME from the CACTUS database to be associated with this particular event, starting at 2012-03-05 02:54~UT, which matches well with the expected launch times from the fixed-phi (02:06~UT), self-similar expansion (02:18~UT), and harmonic mean models (02:25~UT) as listed in the HICAT event information\footnote{see https://www.helcats-fp7.eu/catalogues/event\_page.html?id=HCME\_B\_\kern-.14ex\_20120305\_01 .}. 
The LOWCAT catalogue lists properties of this event such as the COR2 CME mean speed, position angle, and width. The automated algorithm associated the COR2 event with a GOES X1.1 flare starting at 2012-03-05 02:30~UT from NOAA AR 11429. Flare magnitude, location, and start, peak, and end times are listed in the catalogue, followed by the SWPC and SMART properties listed in Table~\ref{tbl:smart_properties} as well as observation time and location. 

\begin{figure}[!t]
\centering
\noindent\includegraphics[width=\textwidth]{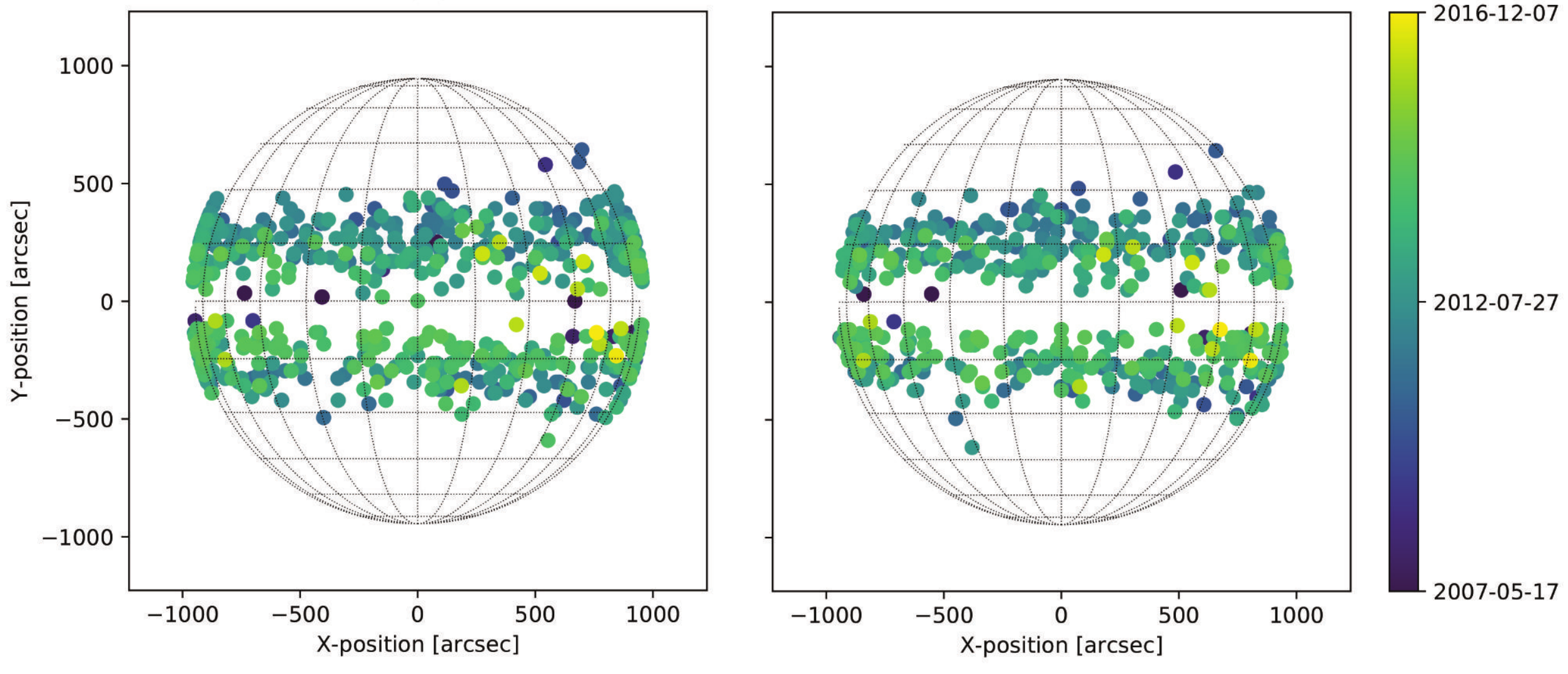}
\caption[Identified flare and SMART locations]{Locations of flare events (left) and SMART detections (right) identified in the LOWCAT catalogue. The colour bar indicates the day of flare peak (left) or magnetogram observation (right).}
\label{fig:flareloc}
\end{figure}

The locations on the solar disk of all flare events and SMART locations in the LOWCAT catalogue are shown in Figure~\ref{fig:flareloc}. The transition of regions towards lower latitudes throughout the solar cycle is clear from the colour bar indicating observation date. A large majority of flare events and their AR sources are from around 2012 - 2014. This is unsurprising considering the period of deep solar minimum around 2009 during which there were considerably less ARs on the solar disk. Note that there is an approximately one year gap in STEREO-A data from 2014 October while the spacecraft was behind the Sun, and data from STEREO-B is not available after 2014 September since contact with the spacecraft had not yet been re-established at the time of the end of the HICAT catalogue development. Comparison between the left and right plots also highlights the difficulty the SMART algorithm has detecting regions on the limb (and magnetic field analysis of these regions would not be as accurate anyway). However, the automated method still identifies $\approx$~45\% of the COR2 CME events to be associated with flare events, and $\approx$~80\% of those flare events have SWPC regions associated with them. 

\subsection{Initial Analysis}\label{s:lowcat_results} 

\begin{figure}[!t]
\centering
\noindent\includegraphics[width=\textwidth]{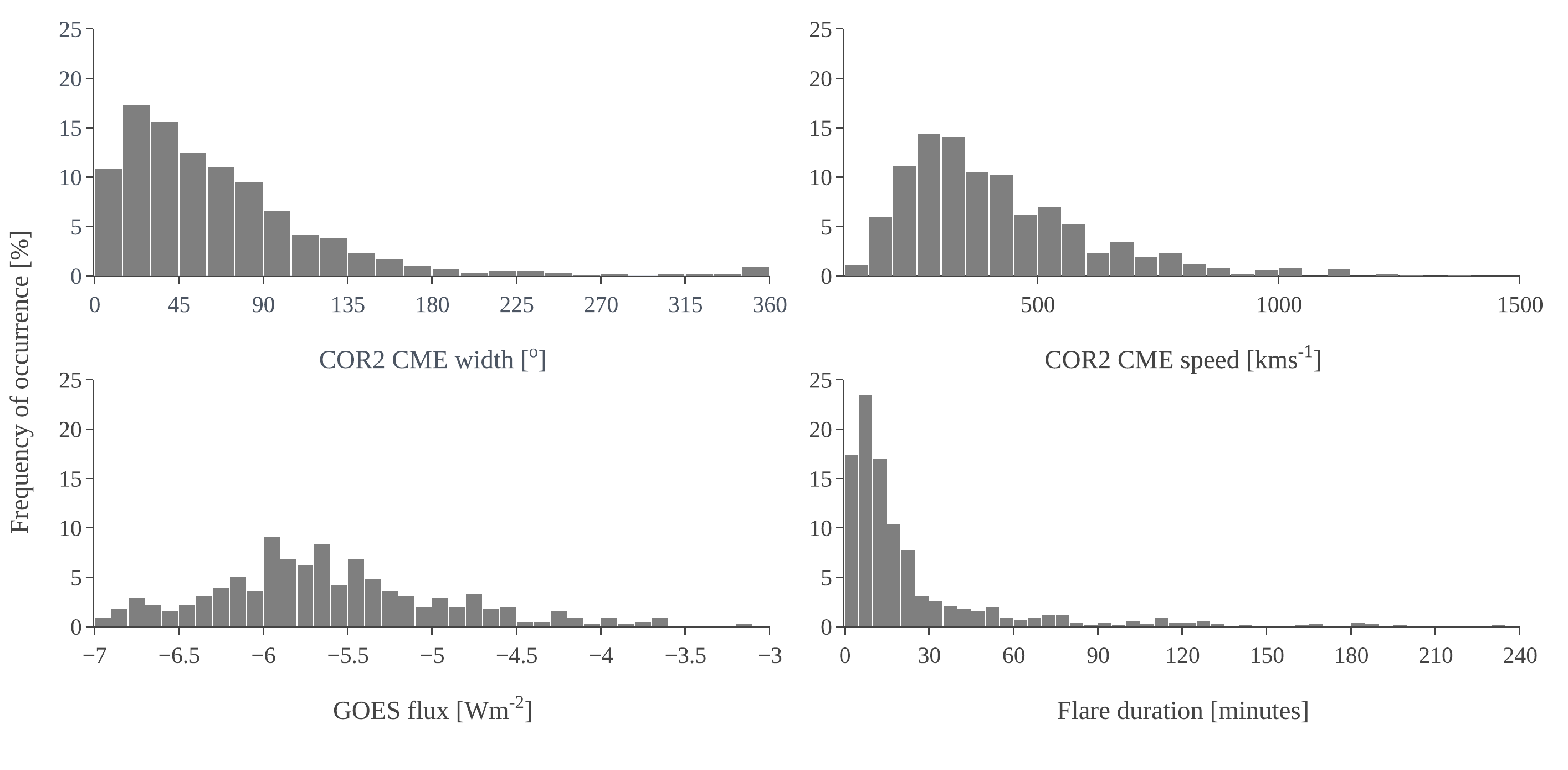}
\caption[Percentage occurrence of CME and flare properties in LOWCAT catalogue]{Percentage occurrence of a number of entries in the LOWCAT catalogue; the upper row shows COR2 CME properties, and lower row SWPC flare event properties. Bins left-to-right top-to-bottom are 15$^{\circ}$, 50 kms$^{-1}$, 0.1 Wm$^{-2}$, and 5 minutes.}
\label{fig:lowcat_cme_flare_histograms}
\end{figure}

\begin{figure}[!t]
\centering
\noindent\includegraphics[width=\textwidth]{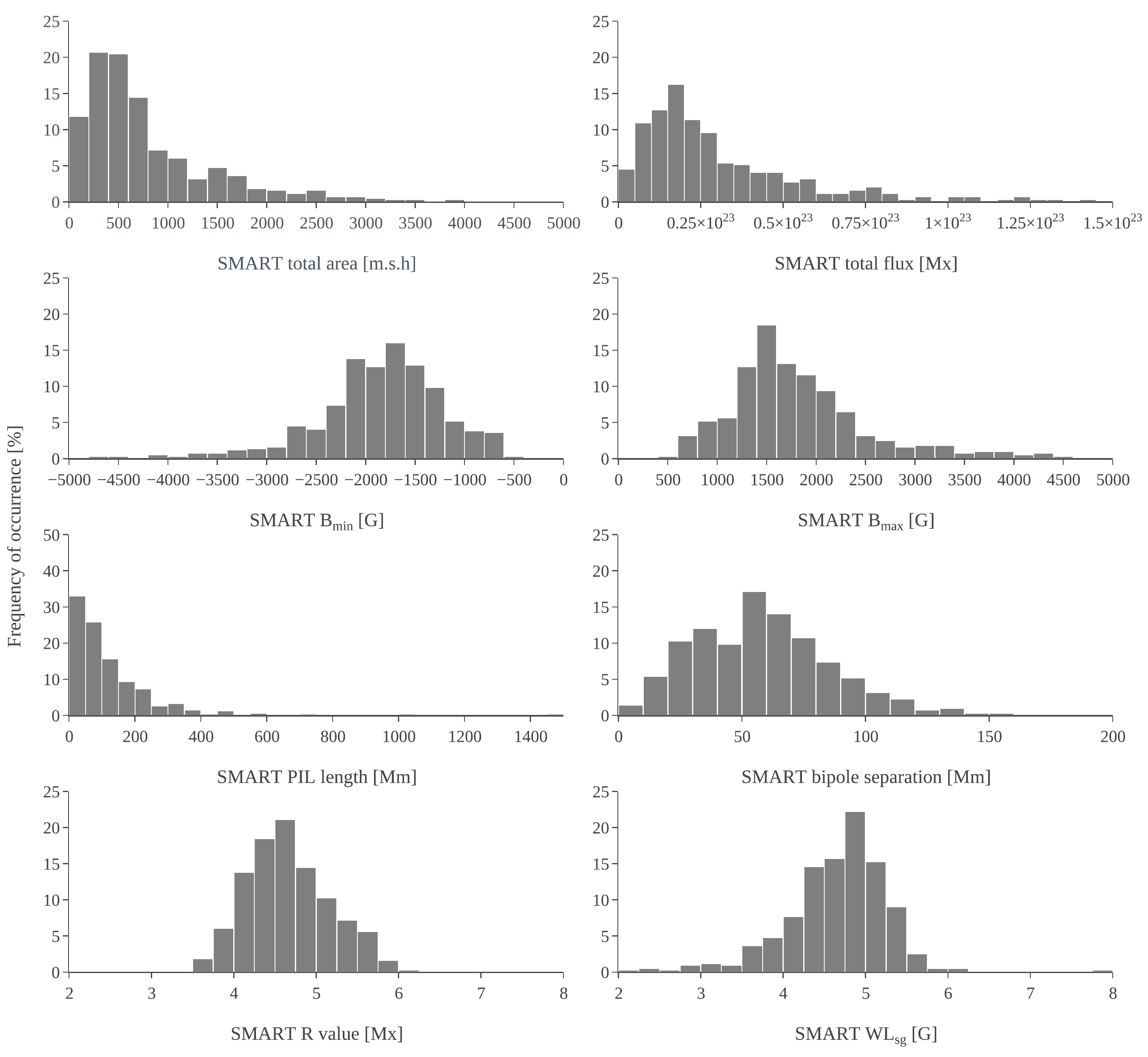}
\caption[Percentage occurrence of of SMART properties in in LOWCAT catalogue]{Percentage occurrence of the SMART properties available in the LOWCAT catalogue. Bins left-to-right top-to-bottom are 200 millionths of a solar hemisphere (m.s.h.), $5\times10^{21}$~Mx, 200 G, 200 G, 50 Mm, 10 Mm, 0.25 Mx, and 0.25 GMm$^{-1}$.}
\label{fig:lowcat_smart_histograms}
\end{figure}

Figures~\ref{fig:lowcat_cme_flare_histograms} and \ref{fig:lowcat_smart_histograms} show the percentage occurrence of all the main properties in the LOWCAT catalogue. A typical range of CME speeds is found as expected from previous studies, confirming the choice of values for the automated algorithm, with a large amount of the CMEs in this catalogue ranging between $\approx$~200~-~600~kms$^{-1}$. The histogram of COR2 CME width shows that there is only a small number of halo events in the catalogue. Most of the SMART properties highlight typical ranges for these AR characteristics, particularly those associated with flaring activity. Interestingly, some parameters are distributed such that most values exist above a certain value, for example the \textit{R} and WL\textsubscript{sg} values. This spurred some initial investigations into potential correlations between the CME, flare, and AR properties in the catalogue.

\begin{figure}[!t]
\centering
\noindent\includegraphics[width=0.9\textwidth]{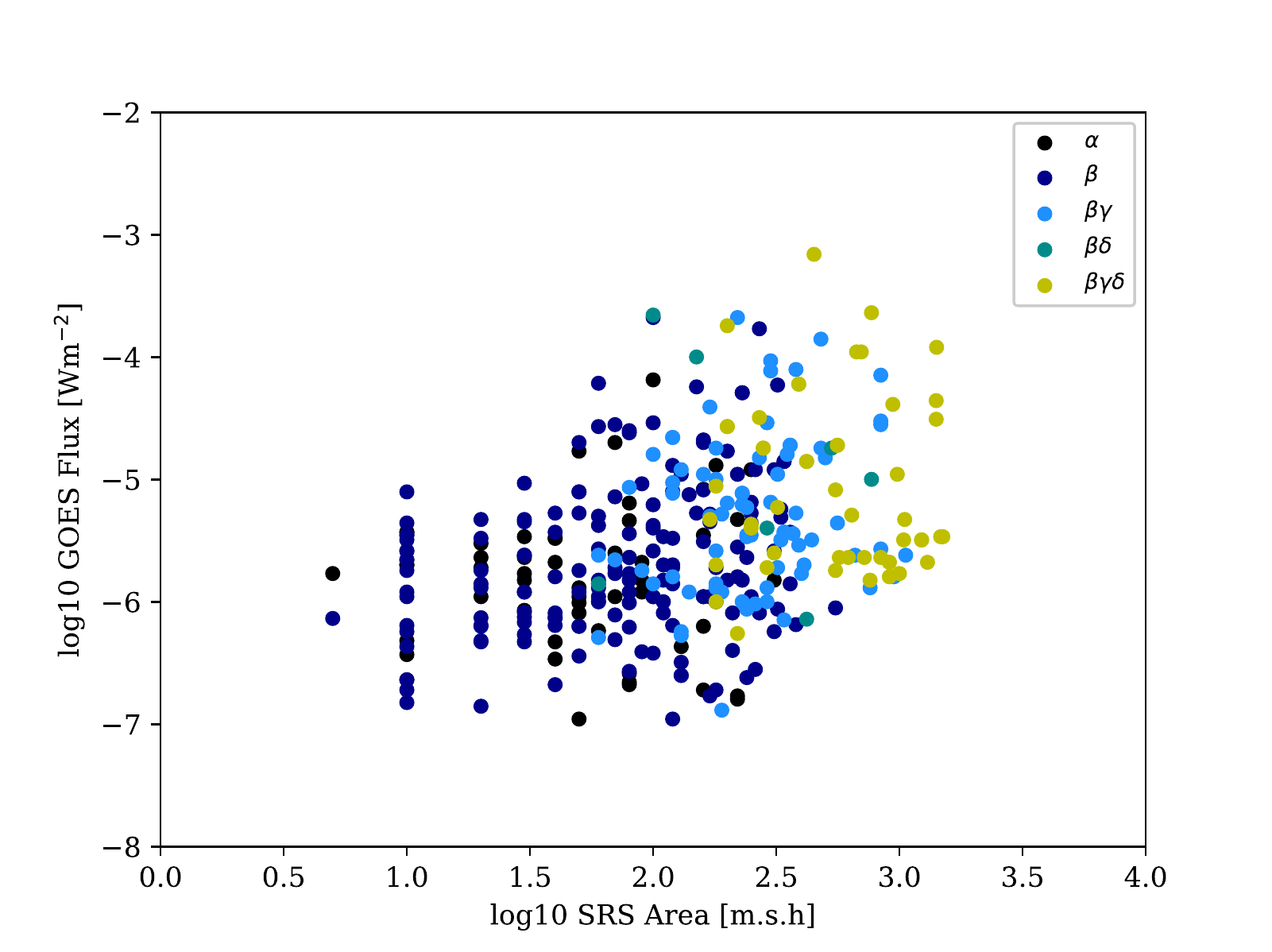}
\caption[GOES flux \textit{versus} NOAA/SRS sunspot group area]{GOES flux \textit{versus} NOAA/SRS sunspot group area, with colours indicating Hale class.}
\label{fig:goes_srs_area}
\end{figure}

In order to confirm regions are being matched by the algorithm as expected (as a supplement to the manual checking), correlations between the flare event and NOAA numbers were examined. It is well established in the research community that large, complex ARs that have a history of flaring are likely to flare again, and AR properties are widely used for flare forecasting purposes \citep[see \textit{e.g.}][and references within]{barnes16}. Figure~\ref{fig:goes_srs_area} shows the peak flare intensity \textit{versus} the area of the LOWCAT regions as identified by SWPC forecasters, with colours indicating Hale class at the nearest time to the flare peak. The plot highlights that bigger flare events in the catalogue occur in bigger regions with more complex Hale magnetic classification sunspot groups, and $\beta\gamma\delta$ regions produce the biggest flares in the catalogue. This is unsurprising, and is particularly reminiscent of Figure 2 of \citet{sammis00}, presenting the same axes and labels for 2,789 NOAA-numbered regions. It must be noted that in the LOWCAT catalogue several events may have the same source region, and the \citeauthor{sammis00} work plots the largest quantities at any time in the AR's lifetime (not necessarily at flare peak). Nevertheless, the expected results encouraged further investigations into other correlations that may be found between various LOWCAT catalogue properties.

\begin{figure}[!t]
\centering
\noindent\includegraphics[width=\textwidth]{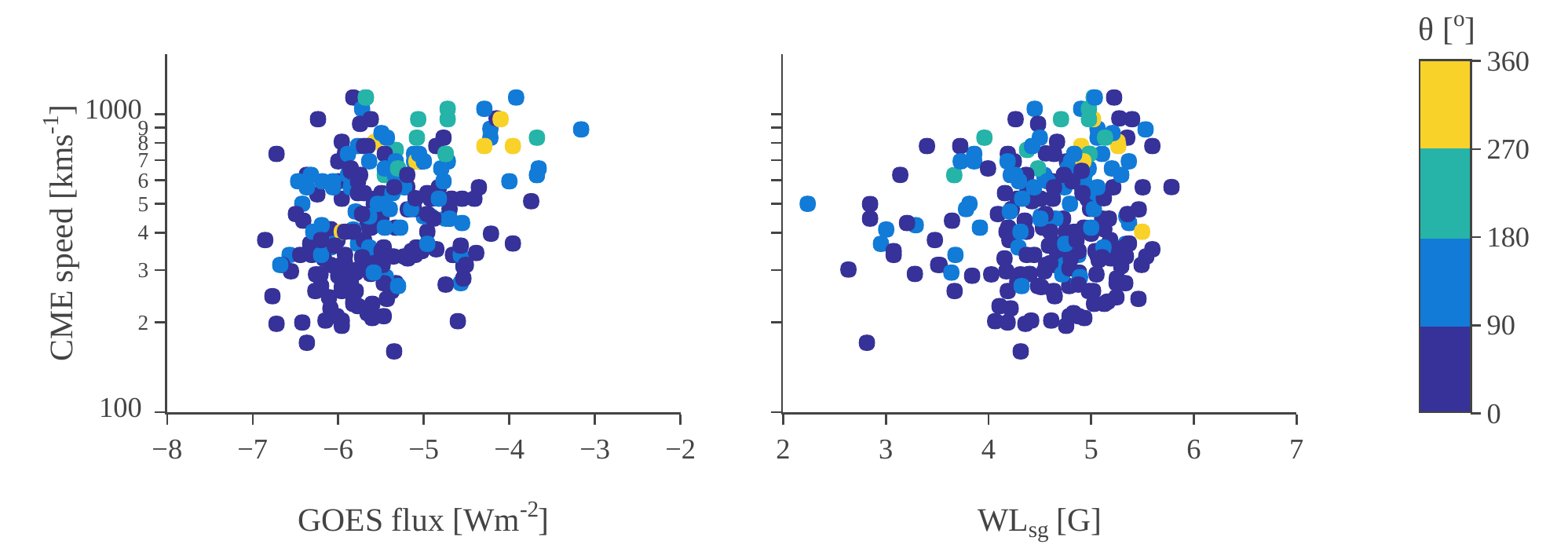}
\noindent\includegraphics[width=\textwidth]{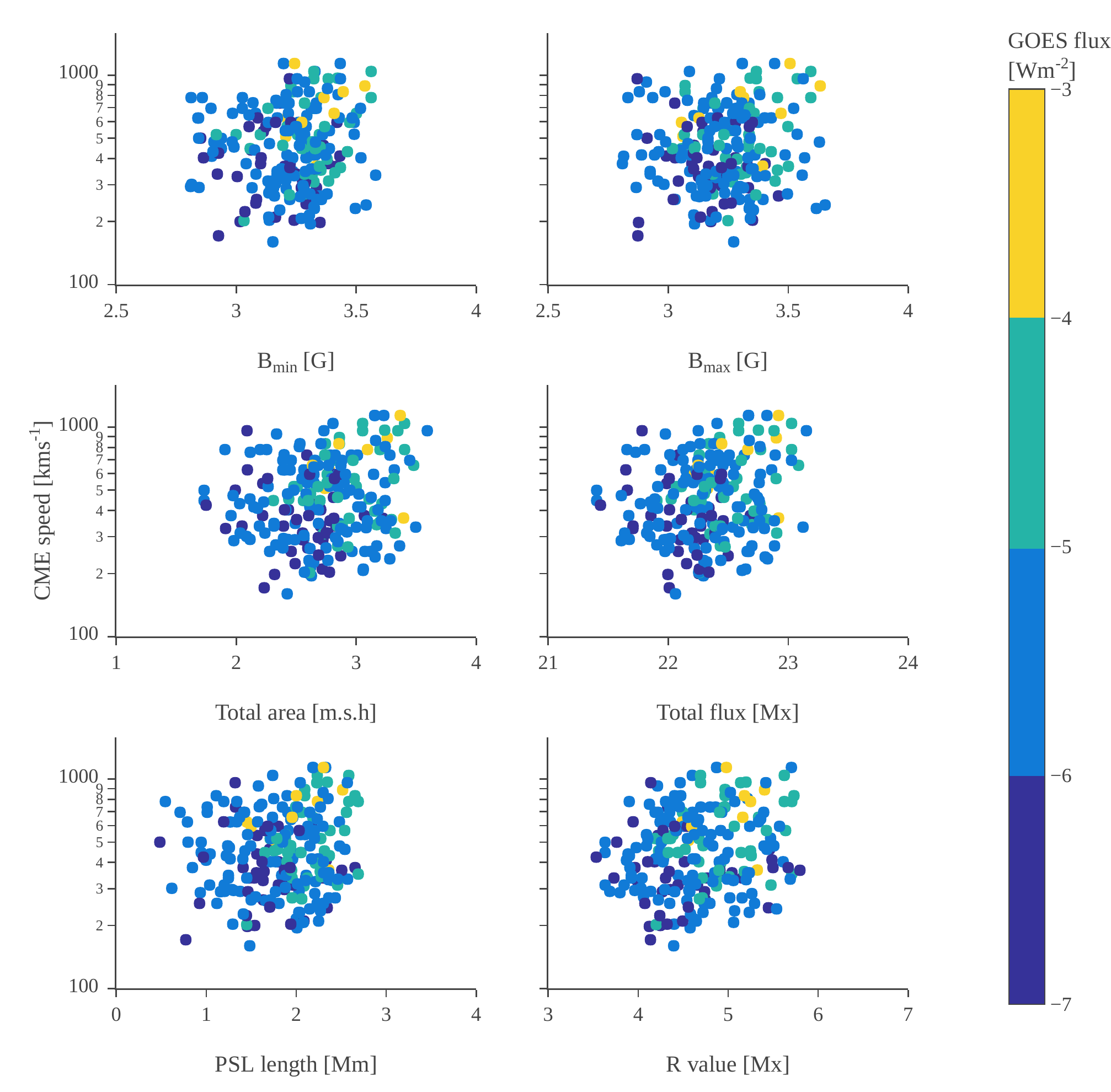}
\caption[Selection of flare and active region properties \textit{versus} CME speed]{Selection of LOWCAT flare and active region properties \textit{versus} CME speed. The upper row colourbar highlights CME width, whilst the colours in the other rows show flare peak intensity (B-class in dark blue, C-class in light blue, M-class in green, and X-class in yellow). Note that all \textit{y}-axes are in log space, and \textit{x}-axes in log10 space.}
\label{fig:cme_flare_smart}
\end{figure}

Figure~\ref{fig:cme_flare_smart} presents comparisons between COR2 CME speeds (\textit{y}-axes) and a selection of other LOWCAT properties (\textit{x}-axes). The upper row in particular shows GOES flare peak intensities (left) and SMART region WL\textsubscript{sg} (right), with colour indicating COR2 CME width, such that yellow highlights halo CME events. It is clear that there is trend of larger flares associated with faster CMEs, although there is some amount of scatter. There is even more scatter found for WL\textsubscript{sg}, however both plots show wider CMEs for larger property values with the halo events tending towards the top right corner of the plot. This suggests the faster wider CMEs are associated with bigger flares and ARs with high WL\textsubscript{sg} values. For example, if forecasters were only interested in CMEs with speeds above 600~kms$^{-1}$, the LOWCAT catalogue suggests flares greater than B class and a WL\textsubscript{sg} value greater than 4 would be of interest. The other rows of Figure~\ref{fig:cme_flare_smart} show more SMART properties, with similar scatter to the upper row, however the colour now highlights the flare magnitude. X-class flares in yellow mainly tend towards the top right corner of the plots similar to the halo events but with more spread in values.

Previous work has suggested thresholds of importance for flare forecasting and CME forecasting alike may exist for these sorts of flaring AR properties \citep[see \textit{e.g.}][]{falconer08, higgins12}. Although one must be careful interpreting these results considering the small number of halo CME and X-class flare events relative to the rest of the catalogue (see Appendix), the results of this initial investigation comparing the LOWCAT CME and AR properties suggests some of these parameters may indeed be useful for CME warning efforts. However, scatter plots can be difficult to interpret, particularly if hoping to draw conclusions regarding what parameters might be most useful for forecasting efforts. To better investigate this idea of warning thresholds, the FLARECAST project database was examined using different analytical techniques to expand upon the AR properties available for comparison.


\section{FLARECAST Active Region Properties}\label{s:flarecast} 

FLARECAST (\url{flarecast.eu} aims to create an automated system for forecasting solar flares with unprecedented accuracy by combining advanced machine learning algorithms with a comprehensive database of flaring activity predictors. The AR property database\footnote{see \url{http://api.flarecast.eu}.} consists of AR photospheric magnetic properties extracted from the Spaceweather HMI Active Region Patches \citep[SHARP,][]{bobra14}. SHARP data products contain AR data maps obtained from SDO/HMI \citep{sdo_hmi} full-disk data. Since the FLARECAST system under development is planned to be operational, all magnetic properties are calculated from the near-real-time (NRT) version of the Cylindrical Equal Area (CEA) de-projection of SHARP data (series hmi.sharp\_cea\_720s\_nrt). See \citet{bobra14} for further details on the SHARP data products. 

Table \ref{tbl:FC_props} lists the AR property groups included in the FLARECAST properties database. Each property group contains several properties. Property groups in Table \ref{tbl:FC_props}, in turn, appear grouped according to the data source they are derived from (SWPC AR catalogues, SHARP vector, LOS magnetograms, and intensity maps). The property database encompasses different AR physical phenomena and areas, for example, magnetic polarity inversion lines, (multi-) fractal distribution of field, photospheric proxies for coronal connectivity, magnetic helicity, and photopsheric flows, among others. The majority of properties included in the FLARECAST property database have been reported as having a relevant association with the occurrence of flares (see references in Table \ref{tbl:FC_props}) and others have been developed and tested in the framework of FLARECAST.

\begin{landscape}
\begin{table}
\caption{FLARECAST Active Region Magnetic Properties}\label{tbl:FC_props}
\begin{tabular}{l l l}     
\hline
Data source		& Property group							& Reference							\\
\hline
SWPC			& Solar Region Summary (SRS) properties		& 									\\
catalogues		& GOES soft X-ray flare events            			& 									\\
\hline
Line-of-sight		& Effective connected magnetic field strength		& \citet{georgoulis07}, 					\\
magnetograms		&										& \citet{georgoulis2010,Georgoulis_2013}		\\
				& Fractal and multifractal parameters			&  \citet{Conlon2008}					\\
				& Fourier and wavelet power spectral indices		& \citet{2015SoPh..290..335G},				\\
				&										& \citet{2002ApJ...577..487A},				\\
				&										& \citet{2008SoPh..248..311H} 				\\
				& Decay index								& \citet{Liu2008}, 						\\ 		                          		     
				&                                                   				& \citet{Zuccarello2014}					\\
				& Magnetic polarity inversion line properties		& \citet{2012ApJ...757...32F},				\\
				&										& \citet{2007ApJ...655L.117S},				\\
				&										& \citet{2010ApJ...723..634M}				\\
				& 3D magnetic null points						& \citet{Haynes2007},					\\
				&										& \citet{Pontin2013}						\\
				& Ising energy								& \citet{Ahmed2010}						\\
				& Magnetic helicity injection rate proxy			& \citet{Park2010},						\\
				&										& \citet{Park2012}						\\
\hline
Vector			& SHARP properties							& \citet{2014SoPh..289.3549B}				\\
magnetograms		& Magnetic helicity injection rate				& \citet{berger_field_1984}				\\
				& Magnetic energy injection rate				& \citet{Kusano2002}						\\
				& Non-neutralized currents					& \citet{Georgoulis2012}					\\
				& Diverging/converging/shear flows				& \citet{Yang2004},						\\
				&										& \citet{Wang2014},						\\
				&										& \citet{Deng2006}						\\
\hline
Intensity images	& Flow field properties						& \citet{Korsos2014}						\\
\hline
\end{tabular}
\end{table}
\end{landscape}

\begin{figure}[!t]
\centering
\noindent\includegraphics[width=0.8\textwidth]{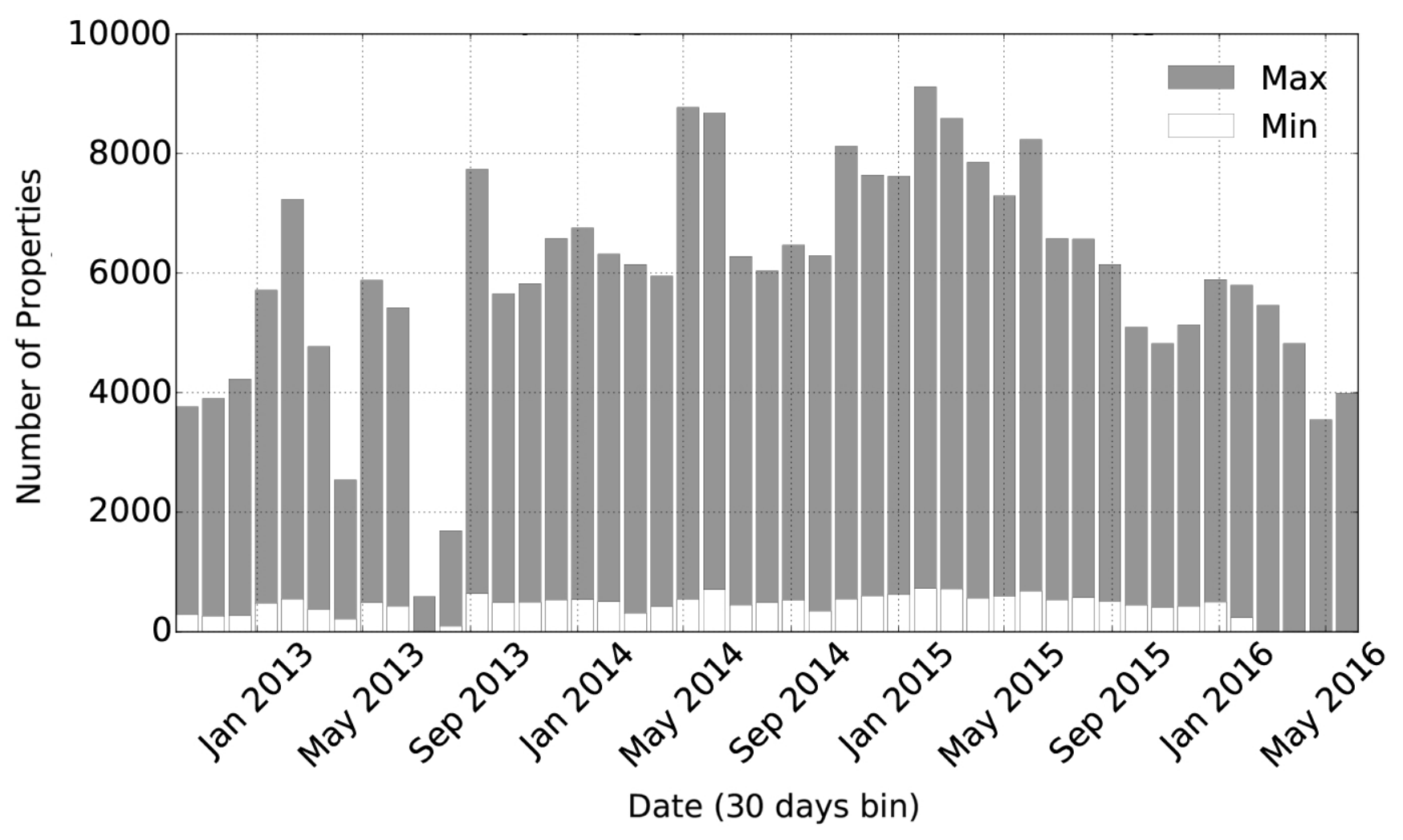}
\caption[FLARECAST active region properties coverage.]{FLARECAST active region property database coverage. Property database covers the entire lifespan of the SHARP data repository (September 2012 - April 2016) with properties calculated every hour. Here the number of property values \textit{per} 30 days is shown. Maximum (grey bars) and minimum (white bars) values for any 30-days bin are a consequence of different properties having longer computation times.}
\label{fig:Fl_data_coverage}
\end{figure}

At present, the FLARECAST property database spans from the beginning of the SHARP data availability (September 2012) to April 2016\footnote{Properties between this date and September 2017 need to be re-calculated due to defective NRT SHARP data.}. Although SHARP data is produced at a 12-minute cadence, properties are calculated only hourly. Figure \ref{fig:Fl_data_coverage} displays the property data coverage. The number of property values present in the database is shown in time with a 30-days bin. Grey and white bars correspond to  maximum and minimal numbers of properties present in each 30-days bin. Minimum and maximum values are due to the differences in computation time among different property-calculating algorithms. For properties that require longer computation times, their values can (at the moment) be found at cadences of 6, 12, and 24 hours. It is found that the maximum number of property values is quite small in June and July of 2013, mainly because many of SHARP data, in the FLARECAST HMI/SHARP service, have insufficient metadata pointing information for the property calculation.

\subsection{Catalogue Comparisons}\label{s:comparison} 

In order to cross-correlate the LOWCAT catalogue to the FLARECAST property database, an algorithm was developed that matches the source regions reported in the former to the regions listed in the latter. Source regions in the LOWCAT catalogue are labelled by their NOAA/SWPC AR number, if available, and/or SMART solar disk location. Regions in the FLARECAST database are primarily identified by HARP number, location, and NOAA number, if available. For each event in the LOWCAT catalogue registered after 1 September 2012, the algorithm searches for all entries in the FLARECAST database with time closest to the associated flare start or peak time, rounded to the hour before. Subsequently, the region matching is done by using the NOAA number (if available) or position. In the case of matching regions by position, a tolerance of 15 degrees in total position ($=[\rm longitude^2 + \rm latitude^2]^{1/2}$) is used. This implies that for a LOWCAT source region to be matched to a FLARECAST region, the former must be located within a circle of radius 15 degrees centred at the latter location. Once the LOWCAT source region has been matched to the FLARECAST region, any photospheric magnetic field property present in the database can then be associated to that LOWCAT event.

\begin{figure}[!t]
\centering
\noindent\includegraphics[width=\textwidth]{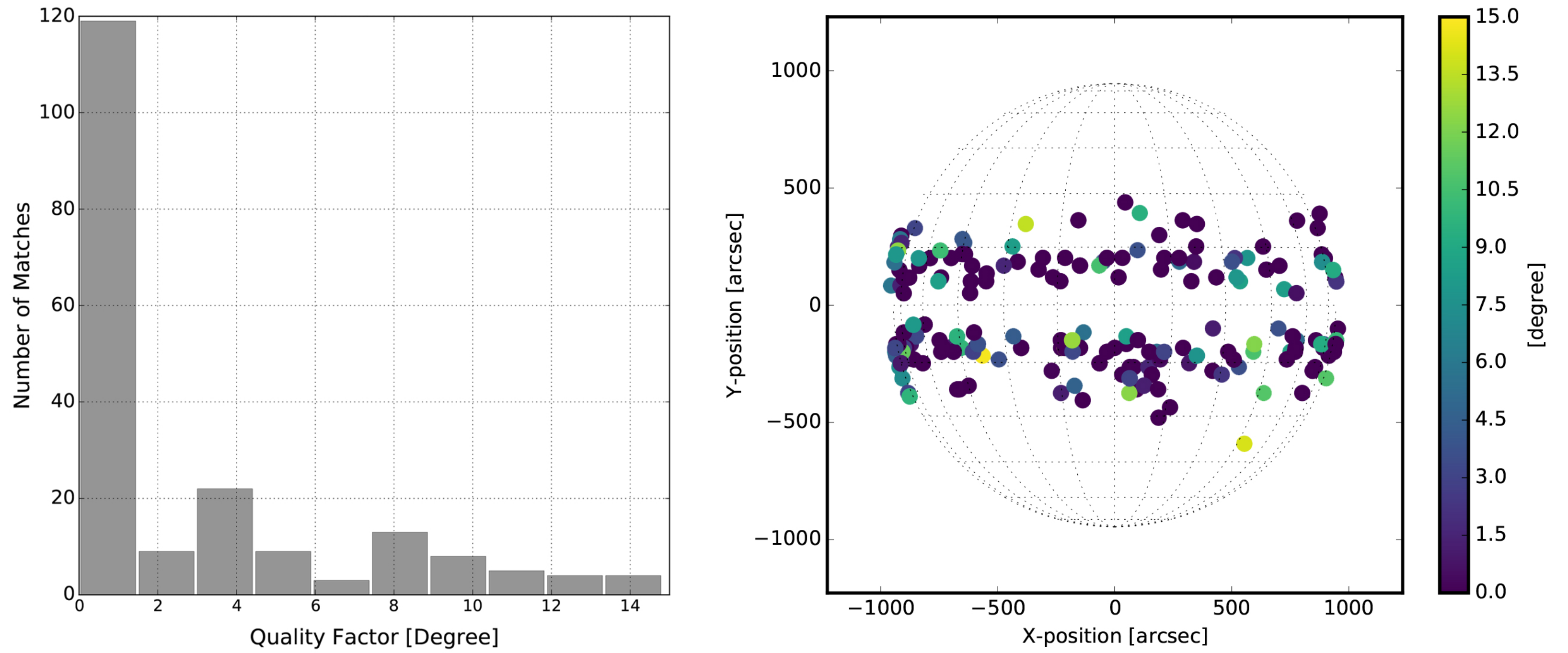}
\caption[Region matching quality]{Quantitative measure for the location-based matching between LOWCAT CME source regions and FLARECAST property database regions. Left and right panels show the histogram and heliographic-spatial distribution of the quality factor. Majority of regions (81\%) are matched within a 5-degree radius from each other and there is no apparent systematic bias with the heliographic location of regions.}
\label{fig:quality}
\end{figure}

It is important to consider that SHARP regions can on some occasions include more than one NOAA region. Such cases are excluded from our analysis since properties can then be overestimated due to the presence of multiple ARs in the field of view. Between 1 September 2012 and 31 April 2016 a total of 812 HI and 665 COR2 CME events are registered in the LOWCAT catalogue. A total of 196 events are matched to FLARECAST regions, according to the criteria stated above. For each LOWCAT source region matched to a FLARECAST region, a matching quality factor is assigned. If the matching was done by means of the NOAA number, a zero value is assigned. On the other hand, if the source region is matched by location, the difference in distance between the regions is used as quality factor. he lower the quality factor the better, with zero values meaning the matching of regions was done exactly. Figure \ref{fig:quality} shows the statistics of region matching quality. On the left, a histogram of the quality factor is displayed, while on the right the quality factor is displayed as a function of the heliographic (HG) position of the LOWCAT source regions. More than half ($\approx$~120 or 60\%) of matched regions are done so exactly or within 1 degree (Figure \ref{fig:quality}, left), although, 90\% of the regions are matched with a quality factor less than ten degrees. On the other hand, no apparent systematic bias of the quality factor with the HG location is seen in Figure \ref{fig:quality}.

\begin{figure}[!t]
\centering
\noindent\includegraphics[width=\textwidth]{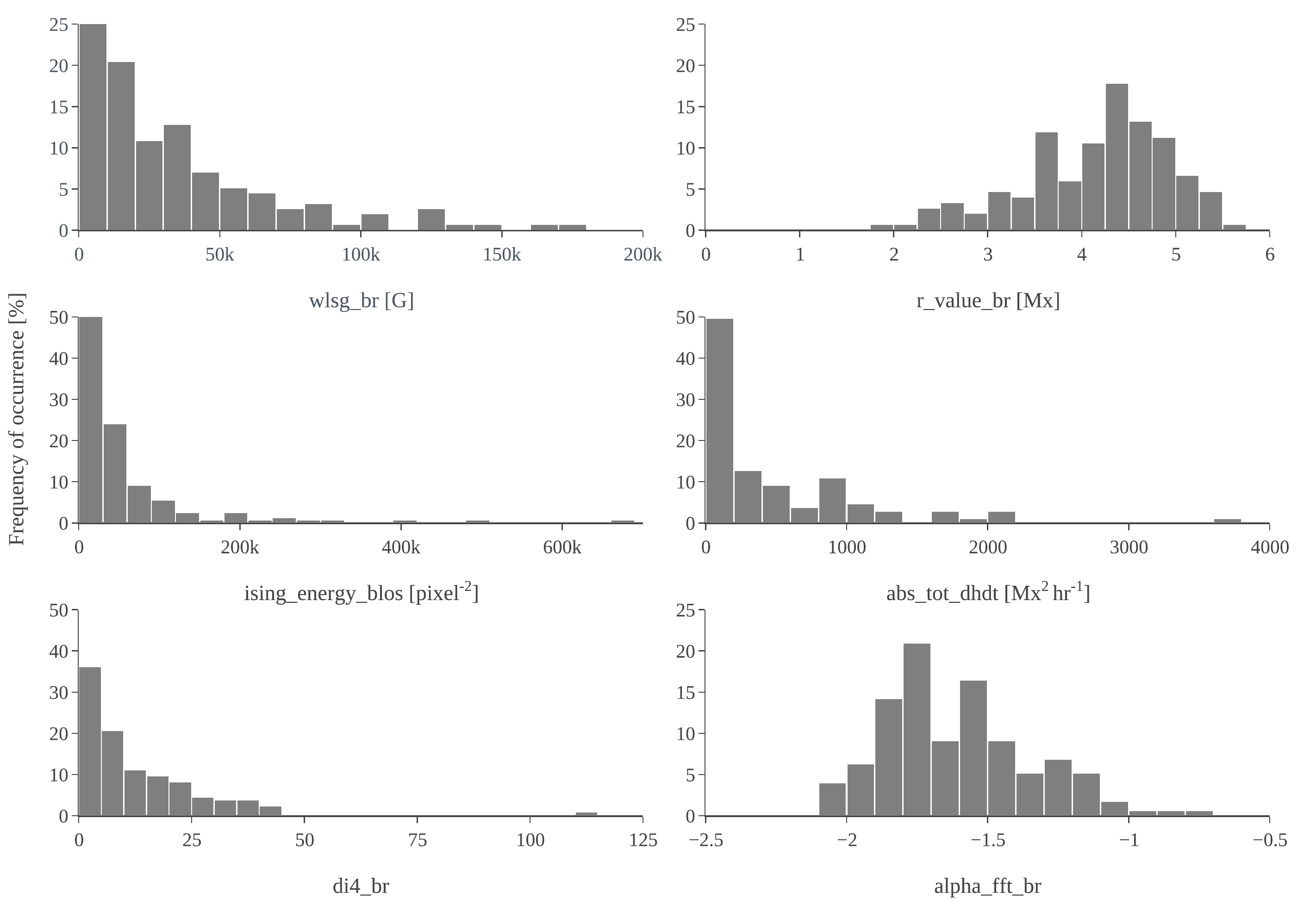}
\caption[Percentage occurrence of properties in the FLARECAST database]{Percentage occurrence of the FLARECAST database active region properties used to compare with the LOWCAT catalogue. Bins left-to-right top-to-bottom are 10,000, 0.25, 30,000, 200, 5, and 0.1.}
\label{fig:flarecast_histograms}
\end{figure}

\begin{figure}[!t]
\centering
\noindent\includegraphics[width=\textwidth]{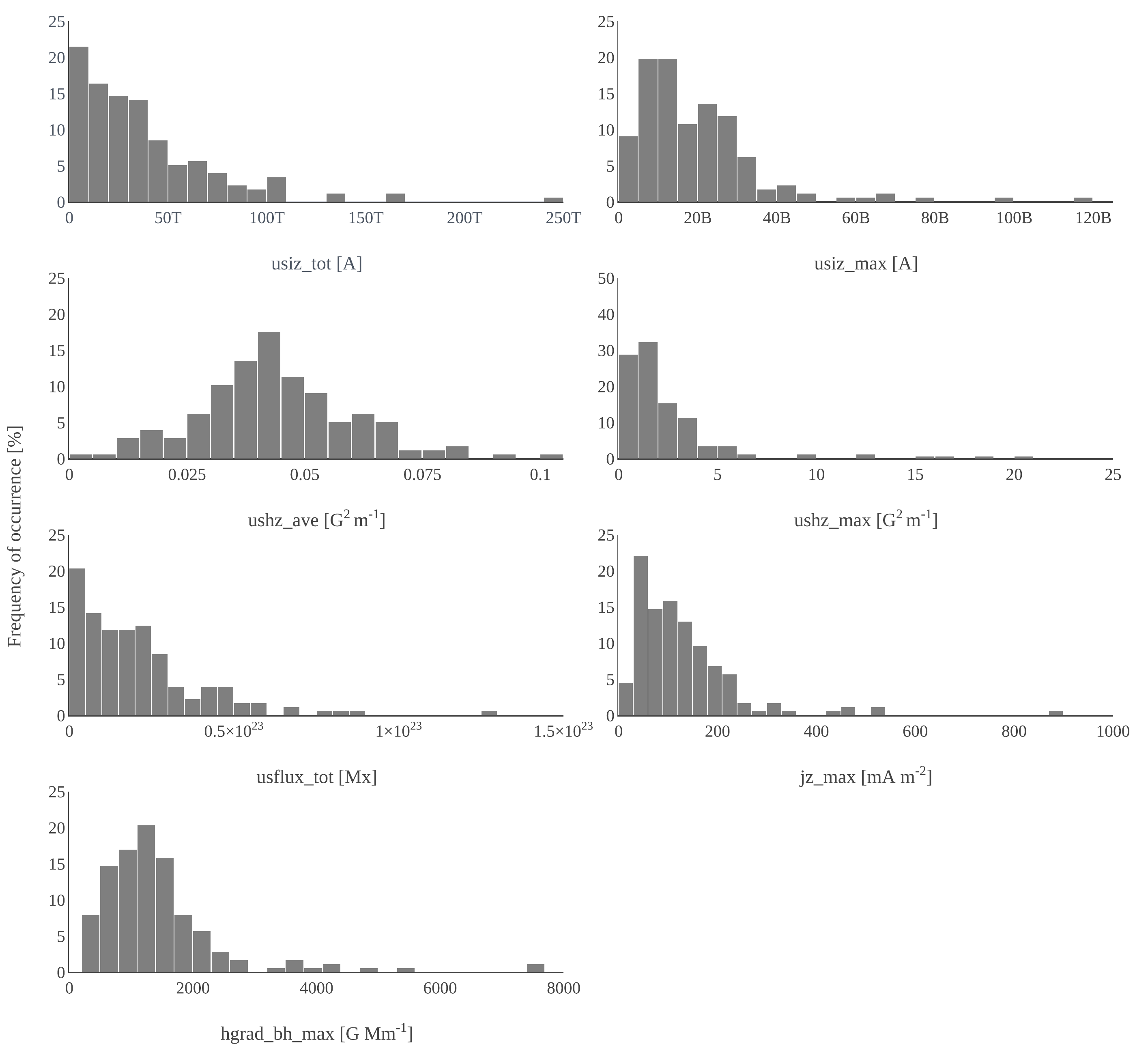}
\caption[Percentage occurrence of properties in the FLARECAST database]{Percentage occurrence of the FLARECAST database active region properties, particularly SHARP properties, used to compare with the LOWCAT catalogue. Bins left-to-right top-to-bottom are $10^{11}$~A, $5\times10^{9}$~A, 0.005~G$^{2}$m$^{-1}$, 1~G$^{2}$m$^{-1}$, $5\times10^{21}$~Mx, 30~mAm$^{-2}$, and 300~GMm$^{-1}$.}
\label{fig:flarecast_sharp_histograms}
\end{figure}

Using the FLARECAST property database allows the investigation of a great variety of photospheric properties in relation to CME production. Of particular interest are properties that are extracted from vector magnetic field data, since only LOS magnetograms are analysed by SMART for the LOWCAT catalogue. Due to the large amount of the properties in the FLARECAST property database (over 100 at the time of publication), this investigation focused on a subset of properties which have been found to be the top predictors for flare occurrence, tested using four different machine learning methods for prediction \citep{CristinaUNIGE}. Table \ref{tbl:props_top15} lists and describes the subset of studied properties selected. The resulting distribution of values for these properties in Table \ref{tbl:props_top15} are shown in Figures~\ref{fig:flarecast_histograms} and \ref{fig:flarecast_sharp_histograms}.

\begin{table}
\caption{Photospheric active-region properties. This table briefly describe the properties from the FLARECAST property database studied in this investigation. Keywords are used throughout as short names for properties.}
\label{tbl:props_top15}
\setlength\tabcolsep{2.5pt}
\begin{tabular}{l l l l}     
\hline
Property 					& Description		& Units		& Keyword \\
\hline
$I_{z,\rm{tot}}$ 				& Total unsigned vertical current ($ = \sum |J_{z}|dA$)  											& A 						& $\textsf {usiz\_tot}$			\\
$I_{z,\rm{max}}$ 			& Max unsigned vertical current ($ = \textrm{max}\{|J_{z}|dA|\}$) 									& A 						& $\textsf {usiz\_max}$ 			\\
WL$_{\rm SG}(B_{r})$  		& Gradient-weighted integral length of neutral line 												& G 						& $\textsf {wlsg\_br}$			\\
$R(B_{r})$ 				& Schrijver's $R$ value from $B_{r}$ 															& G 						& $\textsf {r\_values\_br}$ 		\\
$\Phi_{\rm{tot}}$ 			& Total unsigned flux density ($ = \sum |B_{z}|dA$) 												& Mx 					& $\textsf {usflux\_total}$ 			\\
$H_{c,\rm{ave}}$ 			& Average unsigned current helicity,  														& G$^{2}$ m$^{-1}$ 			& $\textsf {ushz\_ave}$ 			\\
 						& $B_{z}$ contribution ($ = \frac{1}{N}\sum |B_{z}J_{z}|$) 											&						&  							\\
$H_{c,\rm{tot}}$ 			& Total unsigned current helicity, $B_{z}$ contribution 											&  G$^{2}$ m$^{-1}$ 			& $\textsf {ushz\_tot}$ 			\\
 						& $B_{z}$ contribution ($ = \sum |B_{z}J_{z}|$) 													&						& 							\\
$H_{c,\rm{max}}$ 			& Max unsigned current helicity, $B_{z}$ contribution 											& G$^{2}$ m$^{-1}$ 			& $\textsf{ushz\_max}$ 			\\
 						& $B_{z}$ contribution ($ =  \textrm{max}\{|B_{z}J_{z}|\}$) 											& 						& 							\\
$E_{\rm{Ising}}(B_{\rm los})$ 	& Ising energy from $B_{\rm los}$ 															& pixel$^{-2}$ 				& $\textsf {ising\_energy\_blos}$ 	\\
                              			& ($=-\sum_{ij} \frac{S_{i}(B_{\rm los})S_{j}(B_{\rm los})}{d^{2}}$)  										&  						&  							\\
$H_{\rm m}$ 				& Magnetic helicity injection energy 															& Mx$^{2}$ hr$^{-1}$ 		& $\textsf{abs\_tot\_dhdt}$ 		\\
$(L/h_{\rm min})_{\rm tot}$ 	& Sum of PIL-length over minimal height of  													& 						& $\textsf{di4\_br}$				\\
  						& critical decay index ratios 																& 						&							\\
$J_{z,\rm{max}}$ 			& Max vertical current density ($\propto \{ \frac{\partial B_{y}}{\partial x} - \frac{\partial B_{x}}{\partial y}\}$)	&  mA m$^{-2}$ 			& $\textsf {jz\_max}$				\\
$\nabla_{h}(B_{h})_{\rm max}$	& Max horizontal gradient of horizontal field 													&  G Mm$^{-1}$ 			& $\textsf {hgrad\_bh\_max}$ 		\\
$\alpha(B_{r})$ 				& Fourier power spectral index 															& 						& $\textsf {alpha\_fft\_br}$ 		\\
\hline
\end{tabular}
\end{table}

\subsection{Superposed Epoch Analysis}\label{TS_ARP}

This section presents an example of how the FLARECAST property database can be used along with the LOWCAT catalogue for understanding photospheric conditions leading to eruptive flares. One powerful resource of the FLARECAST property database is that time series of AR properties can be constructed to observe the evolution of a property before the flare-CME event. Due to the relatively high number of CME events matched to AR properties, empirical average evolution of properties can be determined by performing a superposed epoch analysis.

\begin{figure}[!t]
\centering
\noindent\includegraphics[width=\textwidth]{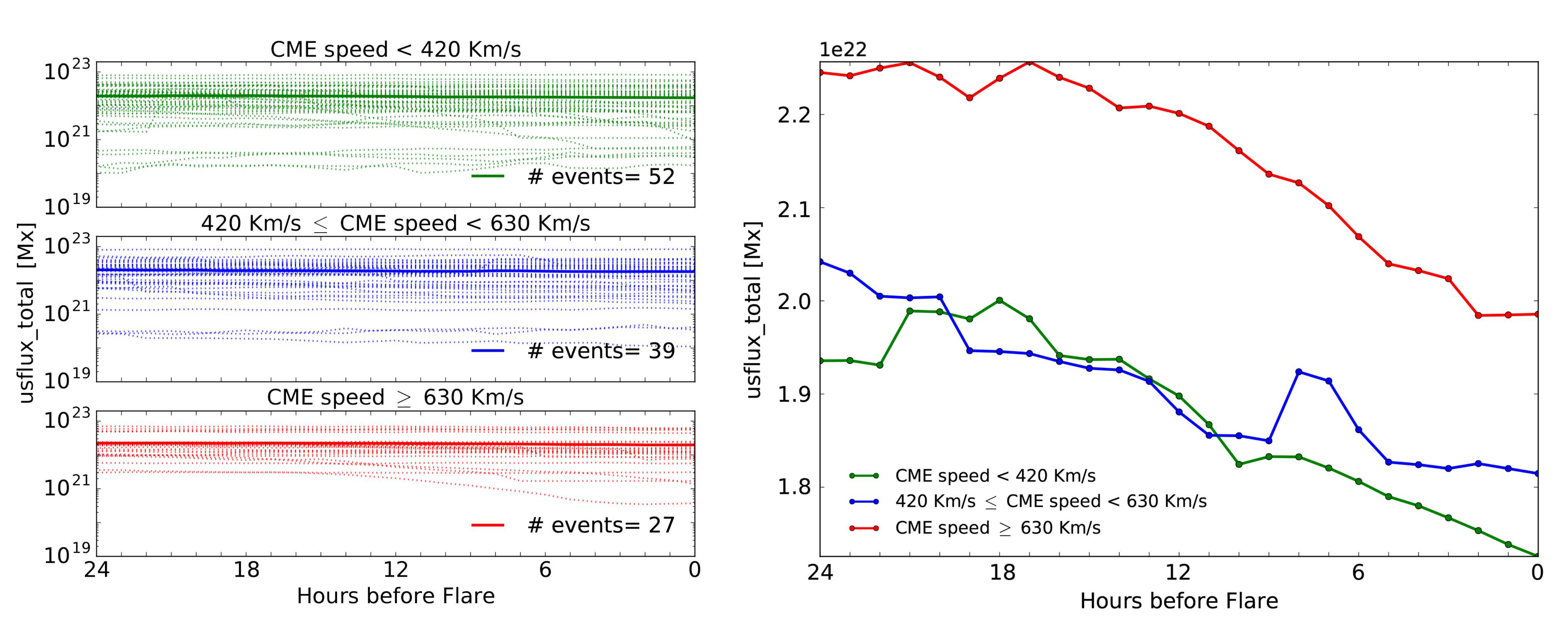}
\caption[Superposed epoch analysis for \textsf{usflux\_total}]{Superposed epoch analysis for the total unsigned magnetic flux ($\Phi_{\rm tot}$, \textsf{usflux\_total}). In the left plot, three panels show the superposed time series for each CME-speed group: slow (top, green), medium (middle, blue), and fast (bottom, red). The plot on the right displays the average time series for each group in the left panels. Included time series and averages are determined to have statistical significance according to the \textit{F} test.}
\label{fig:epoch_analysis_usflux}
\end{figure}

\begin{figure}[!t]
\centering
\noindent\includegraphics[width=\textwidth]{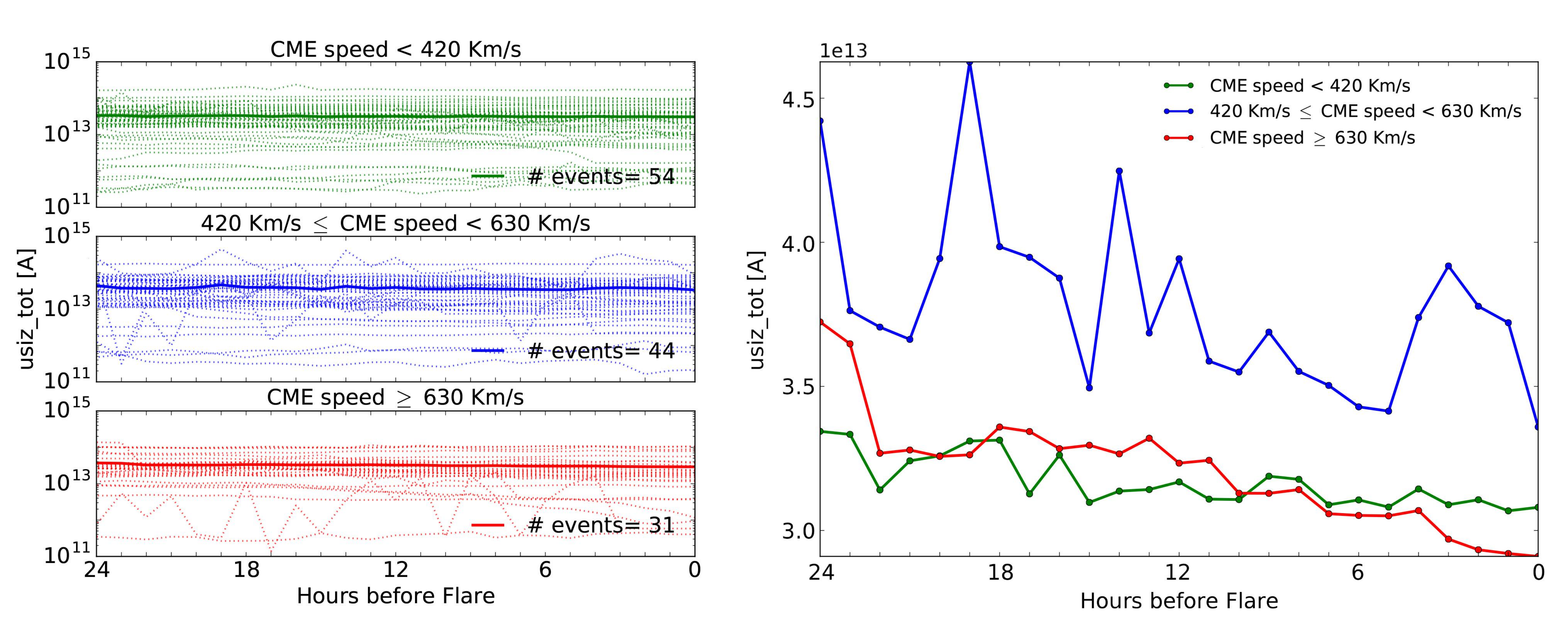}
\caption[Superposed epoch analysis for \textsf{usiz\_tot}]{Same as Figure \ref{fig:epoch_analysis_usflux} but for total unsigned vertical current ($I_{z,\rm tot}$, \textsf{usiz\_tot}).}
\label{fig:epoch_analysis_usiz}
\end{figure}

\begin{figure}[!t]
\centering
\noindent\includegraphics[width=\textwidth]{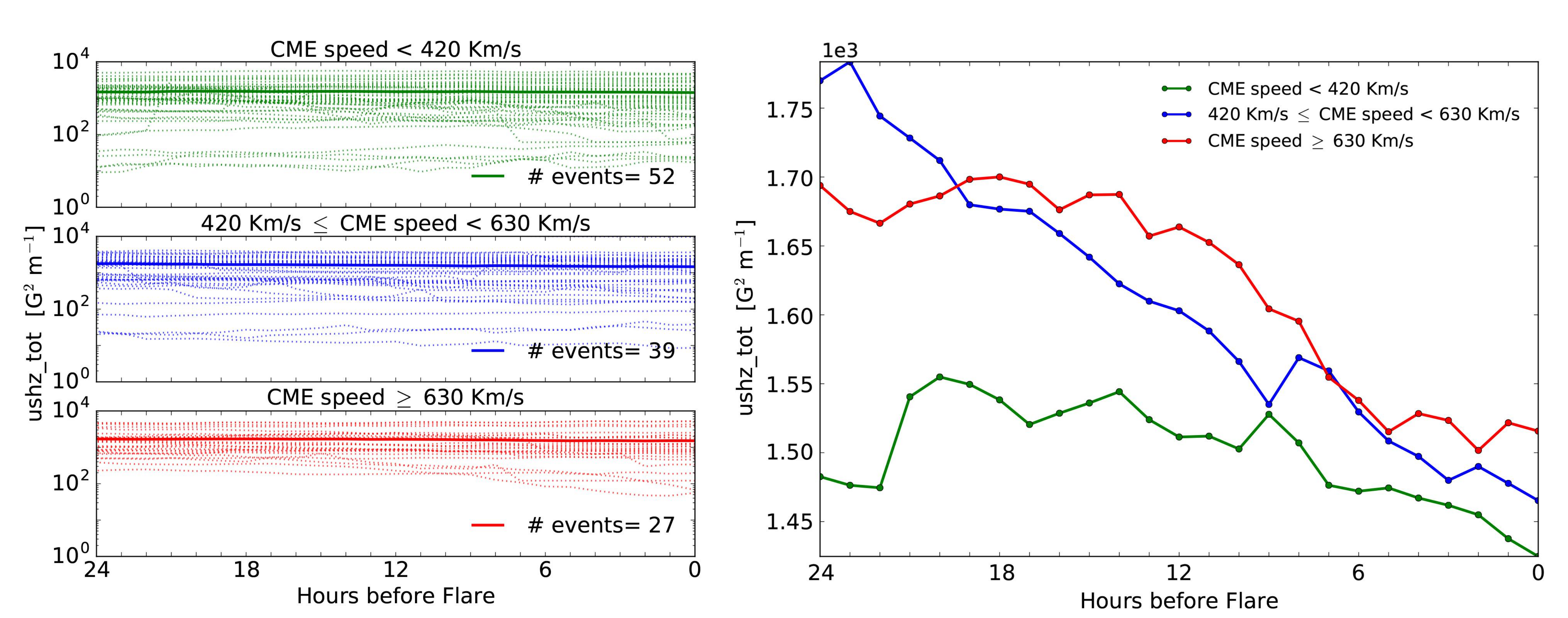}
\caption[Superposed epoch analysis for \textsf{ushz\_tot}]{Same as Figure \ref{fig:epoch_analysis_usflux} and \ref{fig:epoch_analysis_ushz} but for total unsigned $B_{z}$-contribution current helicity ($H_{c,\rm tot}$, \textsf{ushz\_tot}).}
\label{fig:epoch_analysis_ushz}
\end{figure}

For all 196 LOWCAT events matched to FLARECAST regions, the 24-hour evolution prior to the eruption start time is saved for each studied property. No substantial difference was found using the peak instead of start time, since the cadence at which properties are registered is an hour and the difference between start and peak time in a solar soft-Xray flare is usually few to tens of minutes \citep{drake71, 2011LRSP....8....6S}. The superposed epoch analysis is performed separately for three groups according to the CME speed value: i) slow ($v_{\rm CME} <$ 420 km$^{-1}$), ii) average-speed (420 km$^{-1}$ $\leq v_{\rm CME} <$ 630 km$^{-1}$), and iii) fast ($v_{\rm CME} \geq$ 630 km$^{-1}$). These boundary CME-speed values correspond to median($v_{\rm CME}$) and median($v_{\rm CME}) + \sigma(v_{\rm CME})$ values derived from the CME speed distribution (Figure \ref{fig:lowcat_cme_flare_histograms} in Appendix section). In all related plots (\textit{e.g.} Figures \ref{fig:epoch_analysis_usflux}, \ref{fig:epoch_analysis_usiz}, and \ref{fig:epoch_analysis_ushz}), green, blue, and red curves are associated to the slow, average-speed, and fast CMEs groups, correspondingly.

For this analysis only events with $v_{\rm CME} \neq 0$ are considered from the HELCATS catalogue. These zero-speed events correspond to events that have an initial HI-identified CME but no associated COR2 events are found. Out of 196 matched CME-AR cases, 150 have non-zero CME speed. From these cases, 67 are slow CMEs, 50 average-speed CMEs, and 33 fast CMEs. It must be noted that not all cases are included in each CME-speed group, depending on the specific AR property in consideration. This variation is caused by incompleteness of the property time series -- in some cases not all 24 hourly values are present in the property database, and therefore null values can be present. In order to make the present analysis and its results meaningful, a filtering of the time series is performed. In this investigation a similar procedure to \citet{2006JASTP..68..803S} is followed, which suggests the \textit{F} test can be used to determine the statistical significance of the average time series. In this test, the sample is assumed to have a \textit{F}-distribution, and the null hypothesis is rejected at a particular level of confidence. See \citet{2006JASTP..68..803S} and references therein for more details.

\begin{table}
\caption{Photospheric properties showing statistically significant average temporal behaviour from 24 hours prior to CME-associated flares. Properties are listed according to their results in the F-test. From top to bottom the number of significant average time series \textit{per} property \textit{per} CME-speed groups decreases.}
\label{tbl:props_ftest}
\begin{tabular}{l l c c l}     
\hline
Property 		& CME-speed group		& F		& \# of curves & F-test result \\
\hline
$\textsf {usflux\_total}$ & Slow    & 2.44 & 52 & Significant \\
                          & Medium  & 1.36 & 39 & Significant \\
                          & Fast    & 4.99 & 32 & Significant \\
$\textsf {usiz\_tot}$ & Slow    & 1.23 & 54 & Insignificant \\
                      & Medium  & 1.36 & 44 & Significant \\
                      & Fast    & 1.99 & 33 & Significant \\
$\textsf {ushz\_tot}$ & Slow    & 0.85 & 52 & Insignificant \\
                      & Medium  & 4.82 & 39 & Significant \\
                      & Fast    & 2.54 & 33 & Significant \\
$\textsf {wlsg\_br}$  & Slow    & 1.28 & 52 & Significant \\
                      & Medium  & 1.77 & 39 & Significant \\
                      & Fast    & 0.64 & 33 & Insignificant \\
$\textsf {r\_values\_br}$  & Slow    & 1.13 & 54 & Insignificant \\
                           & Medium  & 1.92 & 44 & Significant \\
                           & Fast    & 0.93 & 33 & Insignificant \\
$\textsf {usiz\_max}$  & Slow    & 0.77 & 52 & Insignificant \\
                       & Medium  & 1.70 & 39 & Significant \\
                       & Fast    & 0.91 & 33 & Insignificant \\
$\textsf {alpha\_fft\_br}$  & Slow    & 1.68 & 54 & Significant \\
                            & Medium  & 0.96 & 44 & Insignificant \\
                            & Fast    & 0.61 & 33 & Insignificant \\                    
\hline
\end{tabular}
\end{table}

The average time series depends on the number of samples, $k$, included in each $v_{\rm CME}$ group. In this case, $k$ is determined by the maximum number of null values allowed in any time series, $n_{\rm null}$. As a first approximation to this selection procedure, $n_{\rm null} < 12$, is used. That is, any sample with more than half of their epochs as valid values, is considered. For any null value in the selected time series, interpolated values are calculated. Once $k$ is determined in each CME-speed group for each property, the average time series and its associated \textit{F}-value is calculated. Finally, $F_{\alpha}$, the tabulated \textit{F}-distribution value, is calculated using $n-2$ and $k(n-2)$ degrees of freedom at $\alpha = $ 95\% confidence, where $n=24$ is the number of epochs. For those properties/CME-speed groups where $F > F_{\alpha}$, the average time series is then statistically significant at $\alpha$ level. Table \ref{tbl:props_ftest} summarizes the results from the \textit{F} test, for all properties that showed significant results in at least one CME-speed group.

Figures~\ref{fig:epoch_analysis_usflux}, \ref{fig:epoch_analysis_usiz}, and \ref{fig:epoch_analysis_ushz} show the superposed epoch plots for the top three performing properties from Table \ref{tbl:props_ftest}, namely \textsf{usflux\_total}, \textsf{usiz\_tot}, and \textsf{ushz\_tot}. Figure \ref{fig:epoch_analysis_usflux} (left panels) shows that the initial value of total unsigned flux ($\Phi_{\rm tot}$, \textsf{usflux\_tot}), which is measured 24 hours before the eruption start time, has a dependence on the $v_{\rm CME}$-group. The minimal initial value in each group increases with the CME-speed groups: slow CMEs show a minimal initial value of $\approx~1.0\times 10^{20}$ Mx cm$^{-2}$, medium-speed CMEs show $\approx~1.0\times 10^{21}$ Mx cm$^{-2}$, and fast CMEs show $\approx~3.0\times 10^{21}$ Mx cm$^{-2}$. This finding is consistent with the well-known concept that AR containing large amounts of magnetic flux are prone to eruptive activity. However, and according to Figure \ref{fig:epoch_analysis_usflux}, the higher the total unsigned flux 24 hours before eruption, the bigger the chance of this eruption to be associated to fast CME speeds. On the other hand, Figure \ref{fig:epoch_analysis_usflux} (right) shows that in all three CME-speed groups \textsf{usflux\_total} average time series seem to systematically decrease over the 24-hour period prior to the eruption. This trend could be a manifestation of decaying ARs. Average time series for slow and average-speed CMEs (green and blue) show close values between hours 21 and 9 before eruption, but for any other epoch AR that produced average-speed CMEs show higher $\Phi_{\rm tot}$ values. Values of $\Phi_{\rm tot}$ for fast-CMEs associated ARs (red) are well above the two other categories.

Corresponding superposed epoch analysis for the total unsigned vertical current ($I_{z,\rm tot}$, \textsf{usiz\_tot}) is shown in Figure \ref{fig:epoch_analysis_usiz}. As in the previous case, the minimal initial value of $I_{z,\rm tot}$ seems to increase with the CME-speed group: 2$\times$10$^{11}$ A, 5$\times$10$^{11}$ A, and 5$\times$10$^{12}$ A for slow, average-speed, and fast CMEs, respectively. In comparison to $\Phi_{\rm tot}$, more variations of $I_{z,\rm tot}$ (up to several  order of magnitudes) can be seen in all three panels. This variation could be spurious, a consequence of the interpolation done for null values. Although, their appearance on the average time series (Figure \ref{fig:epoch_analysis_usiz}, right) for average-speed and fast CMEs (the statistically significant cases for this property) suggests that rapidly changing vertical currents could be present in regions before erupting. This effect, real or artefact, could be better studied by constructing higher-cadence time series, particularly making use of the highest SHARP data cadence of 12 min. It is important to remark that statistically significant time series in Figure \ref{fig:epoch_analysis_usiz} (right, blue and red curves) decrease in value with increasing $v_{\rm CME}$. 

Figure \ref{fig:epoch_analysis_ushz} displays the superposed epoch analysis for the total unsigned current helicity due to the $B_{z}$ component ($H_{c,\rm tot}$, \textsf{ushz\_tot}). As in previous cases, time series of $H_{c,\rm tot}$ \textit{per} CME-speed group (Figure \ref{fig:epoch_analysis_ushz}, left) show the spread over the property value's range decreasing from slow to medium-speed to fast CME speeds. In particular, minimal initial values (24 hours before eruption) of total $B_{z}$-contribution unsigned helicity are observed to increase with the CME-speed group: $\lesssim$ 10 G$^{2}$ m$^{-1}$ for slow CMEs, $\approx$~20 G$^{2}$ m$^{-1}$ for medium-speed CMEs, and 200 G$^{2}$ m$^{-1}$ for fast CMEs. The level of hourly variations within the 24-hour study period appear to be in between those of $I_{z,\rm tot}$ and $\Phi_{\rm tot}$. This last finding becomes clearer by inspection of the average time series/CME-speed group (Figure \ref{fig:epoch_analysis_ushz}, right). The average time series for medium-speed CMEs (blue) is also observed to systematically decrease over a day before the eruption. On the other hand, fast and slow CMEs (red and green, respectively) appear to have a seemingly flat period, between hour 21 and 9, then a decrease over two hours, follow by a less steep decrease from 7-6 hours before eruption. Although, only red and blue curves are statistically significant according Table \ref{tbl:props_ftest}. More AR properties could display statistically significant results in the superposed epoch analysis by performing a better filtering of the timeseries sample.

\begin{figure}[!t]
\centering
\noindent\includegraphics[width=\textwidth]{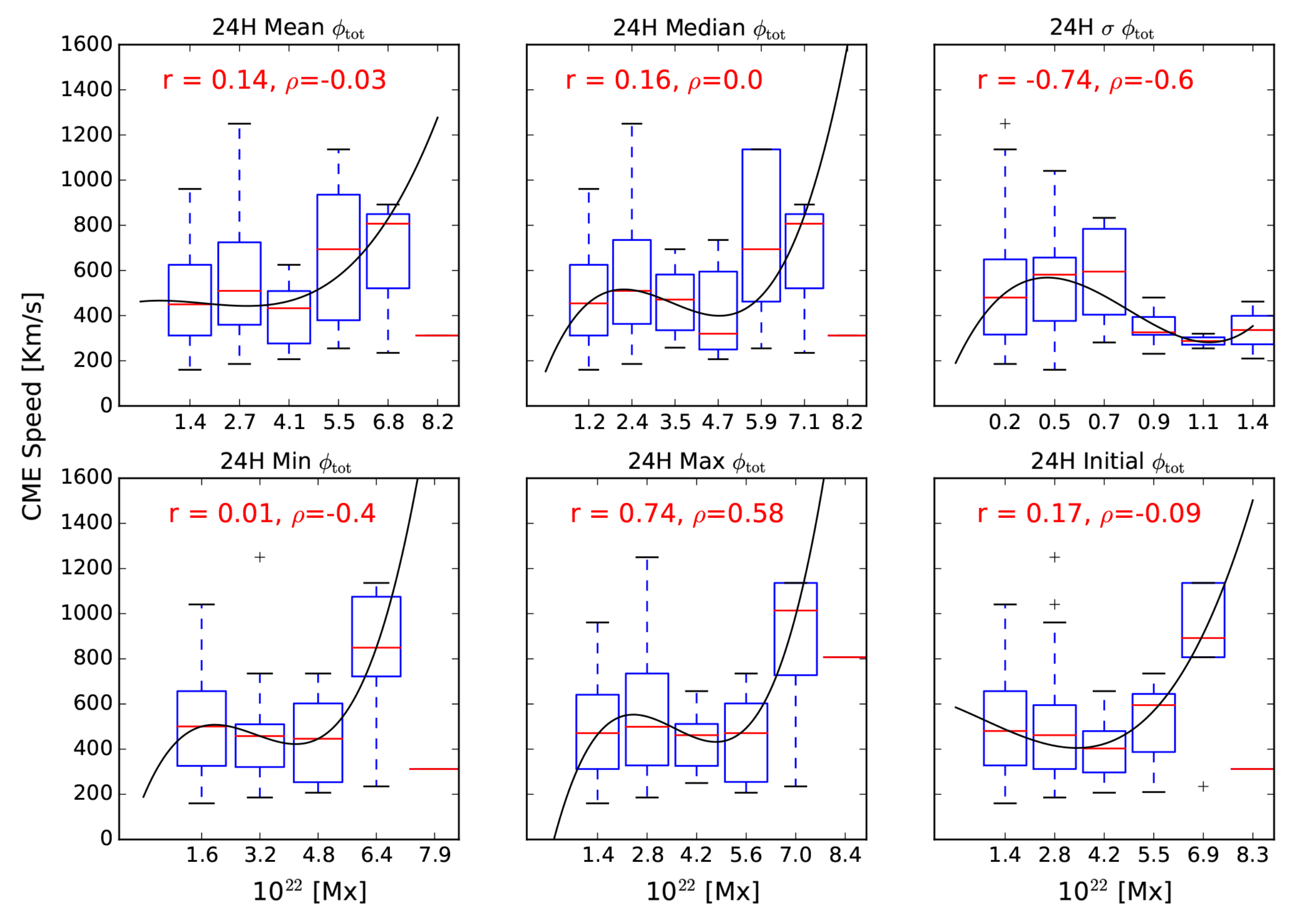}
\caption[Box-and-whiskers plots of CME speed \textit{versus} parametrizations of the 24-hour $\Phi_{\rm tot}$ distribution]{Box-and-whiskers plots of CME speed \textit{versus} the 24-hour $\Phi_{\rm tot}$ mean, median, standard deviation (top panels), minimum, maximum, and initial value (bottom panels). Solid black line corresponds to a fit of the median values (red horizontal lines) using a third-degree polynomial. Values in the horizontal axis are binned so the residual between the data and fit is minimized.}
\label{fig:boxplot_phi_tot}
\end{figure}

Time series of photospheric AR properties associated to eruptive events allows us to investigate statistical associations between CME characteristics (speed and/or angle) with parametrizations of the 24-hour distribution of values. It is clear from Figure \ref{fig:cme_flare_smart} that scatter plots display large levels of dispersion to claim any (linear or non-linear) correlation between the variables. As an alternative, box-and-whisker plots allow a better defined association between CME characteristics and photospheric properties. Figure \ref{fig:boxplot_phi_tot} presents the box-and-whisker plots for CME speed, $v_{\rm CME}$, \textit{versus} 24-hours $\Phi_{\rm tot}$ distribution mean, median, and standard deviation (top row) and minimum, maximum, and initial values (bottom row). Box-and-whisker plots present groups of numerical data through their distribution quartiles. In order to do so, the data is binned in the abscissa variable thus providing a distribution of values in the ordinate variable \textit{per} bin. For a bin in any panel of Figure \ref{fig:boxplot_phi_tot}, the red horizontal bar corresponds to the median value (second quartile), and the top and bottom of the blue box mark the third and first quartile, correspondingly. The whiskers (dashed lines) represent the maximum and minimum values of the distribution, excluding outliers (black crosses). In this way, the CME speed is represented by the median value of the distribution in each $\Phi_{\rm tot}$ bin.

In order to test for a linear or non-linear association between the variables, for each box-and-whiskers plot in Figure \ref{fig:boxplot_phi_tot} a third polynomial curve (solid black line) is estimated by fitting the median values. In addition, linear (Pearson) and non-linear (rank Spearman) correlation coefficients are calculated and displayed. The polynomial degree and number of bins in the abscissa coordinate were selected to minimize the residuals between the binned data and fit. By visual inspection of the data-fit curves, median values of CME speeds seem to show an increasing relation with all 24-hour $\Phi_{\rm tot}$ distribution-related quantities in Figure \ref{fig:boxplot_phi_tot} but one, the 24-hour $\Phi_{\rm tot}$ standard deviation. At least two relationships show moderate to strong correlations (value $s= 0.5$~--~$0.75$): maximum value and standard deviation of the 24-hour $\Phi_{\rm tot}$. In the case of $v_{\rm CME}$ and the 24-hour $\sigma(\Phi_{\rm tot})$ (although the number of data points in each plot is small; between 5 and 7), correlation coefficients show the CME speed having a moderate, negative, and slightly more linear relation with the 24-hour $\Phi_{\rm tot}$ standard deviation. This statistical association might suggest that fast CMEs can originate from regions in which the total unsigned flux varies little over the 24-hour window leading to the eruption. On the other hand, $v_{\rm CME}$ and the maximum value of $\Phi_{\rm tot}$ 24 hours before eruption, show a moderate, positive, and linear correlation association, implying that the higher the maximum measured value of $\Phi_{\rm tot}$, a faster CME could be expected. It should be kept in mind these statistical results correspond to CMEs associated to flares and ARs. Therefore similar analyses using non-eruptive flaring or quiet active regions should be performed for completeness. Such comparative analyses could then provide precursor signals and/or probabilistic relationships for CME characteristics inferred from the evolution of AR photospheric properties.


\section{Conclusions and Future Work}\label{s:concl} 

An automatic back propagation algorithm has been developed to identify source regions associated with CME events. The algorithm traces HI-identified events back to COR-identified events, associates flares with these COR-identified events, and then associates the flares with their AR sources. SMART is used to obtain complex photospheric magnetic field properties of these identified ARs. This algorithm (as summarised in Figure~\ref{fig:flowchart}) has been run on the HELCATS project HICAT catalogue, producing the LOWCAT catalogue. 

The LOWCAT catalogue is available open-access online \citep{murray17}, containing properties of the $\approx$~2,000 HICAT CME events, $\approx$~720 flares ($\approx$~45\% of the CME events), and $\approx$~550 NOAA/SWPC ARs sources. Although some events may be missed due to the strict algorithm constraints, the automated method enables easier study of such a large number of events, unlike the significant time effort involved with manual association. The solar cycle period during which observations were taken also limits associations, however the resulting numbers do compare well with similar large catalogue studies. \citet{andrews03} found all X-class flares and $\approx$~55\% of the M-class flares were associated with 311 LASCO-observed CMEs from 1996 to 1999, supporting the conclusion of \citet{harrison95} that particular flare intensities and durations do not guarantee CME association. \citet{yashiro05} found 20\% C-class flares associated with 1,301 CMEs in a larger statistical study, and half of the CMEs associated with the C-class flares were invisible in LASCO. Regarding eruption location on the solar disk, \citet{lara08} previously found that CME activity has a slight preference for the northern and southern hemispheres during the first and second halves of a solar cycle, respectively. The LOWCAT events seem generally evenly distributed in Figure~\ref{fig:flareloc}, although the data gaps do not allow any more conclusions to be drawn regarding spatial correlations.

The simple correlations between the properties available within LOWCAT presented in Section~\ref{s:lowcat_results} are comparable to previous investigations involving other catalogues of LASCO CME observations (rather than STEREO). \citet{yashiro05} found CMEs associated with X-class flares were faster, with wider extents, than those associated with C-class flares. In a study of 86 flare-associated CMEs, \citet{guo07} found faster flare-associated CMEs tend to be accompanied by more intense flares. Further, for CMEs associated with 55 ARs, \citeauthor{guo07} found that fast CMEs initiated from ARs with large total flux and effective distance (a parameter quantifying magnetic complexity). Similar trends between CME, flare, and AR properties are found in Figure~\ref{fig:cme_flare_smart}, with the fast halo CMEs being associated with X-class flares and big, complex ARs. However, it is apparent that the traditional scatter plot method generally does not clearly highlight potential correlations, and the more detailed analysis used for the LOWCAT-FLARECAST comparisons has proven to be more insightful in this case into what particular AR properties may be most useful for CME warning purposes.

The initial exploration of the FLARECAST AR property database in relation to eruption/CME characteristics demonstrate the importance of investigating full-vector field photospheric properties, in accordance with previous studies \citep{tiwari15,ventakrishnan03}. It is made clear in this investigation that the 24-hour time evolution of photospheric properties provides more insight than single values at the eruption onset and improve the statistical quality of results. According to the results presented here, three properties show significant statistical association with the CME speed: total unsigned flux, total unsigned vertical current, and the total unsigned $B_{z}$-contribution helicity current. In all three cases, the minimal property value 24 hours before eruption showed an increasing tendency from slow, to medium-speed, to fast CMEs. The evolution during the study period showed two interesting behaviours of the properties: for eruption-associated ARs, on average i) the total unsigned magnetic flux decreases relatively steadily, ii) total unsigned vertical current shows short-time oscillations. On the other hand, using alternative data representations to traditional scatter plots, empirical relations between the CME speed and the 24-hour property distribution descriptors for total unsigned magnetic flux were found: i) a negatively linear correlation between CME speed and the standard deviation, and ii) a positively linear correlation between CME speed and the maximum measured values in 24 hours.

The results presented in this paper highlight some initial investigations with the properties listed in the new LOWCAT catalogue and FLARECAST database. It must be noted that the analyses presented here serve to show examples of what can be done with the datasets in the hope they will be used extensively in the future. With both data sets freely available online the authors encourage the community to continue these initial investigations. Comparison with other catalogues may prove interesting in the future, such as associating the CME events with filament eruptions \citep[\textit{e.g.} the SDO catalogue of ][]{mccauley15}. The HELCATS project has also developed numerous other catalogues (such as ARRCAT, CIRCAT, ICMECAT, LINKCAT, and RADCAT) which are all also open access and could be used for more detailed studies of CMEs as they propagate out from their source into interplanetary space (some impacting Earth).

Ultimately these statistical analyses on large datasets provide important insight into the link between CME and flare events, as well as characteristics of eruptive active regions. The correlation algorithms presented here have been written such that other catalogues could be easily integrated, for example they could be run with LASCO CME catalogues in the future. The code is freely available to anyone who may wish to use it, and may prove useful for future operational CME forecasting efforts.


%


%
\begin{acks}
The FLARECAST database is available at \url{http://api.flarecast.eu/}, HICAT catalogue at \url{https://www.helcats-fp7.eu/catalogues/wp2\_cat.html}, and LOWCAT at \url{https://figshare.com/articles/HELCATS\_LOWCAT/4970222}. The authors wish to acknowledge the use of the Overleaf to prepare the manuscript, and also the following Python libraries and packages used when creating the figures in this paper: Astropy, Matplotlib, NumPy, pandas, Plotly, SciPy, and SunPy. The STEREO/SECCHI data used here are produced by an international consortium of the Naval Research Laboratory (USA), Lockheed Martin Solar and Astrophysics Laboratory (USA), NASA Goddard Space Flight Center (USA), Rutherford Appleton Laboratory (UK), University of Birmingham (UK), Max-Planck-Institut f\"{u}r Sonnensystemforschung (Germany), Centre Spatial de Li\`{e}ge (Belgium), Institut d'Optique Th\'{e}orique et Appliqu\'{e} (France), and Institut d’Astrophysique Spatiale (France). SDO is a mission for NASA's Living With a Star (LWS) program, with the SDO/HMI data provided by the Joint Science Operation Center (JSOC). E.C., P.Z., and S.A.M were supported by the European Union Seventh Framework Program under grant agreement No. 606692 (HELCATS project). J.G.A., S.A.M., and S.-H.P were supported by the European Union Horizon 2020 research and innovation program under grant agreement No. 640216 (FLARECAST project). VB acknowledges support of the CGAUSS (Coronagraphic German and US Solar Probe Plus Survey) project for WISPR by the German Space Agency DLR under grant 50 OL 1601. S.A.M. acknowledges the IRC Postdoctoral Fellowship Scheme and AFOSR award FA9550-17-1-039. The authors would like to thank the anonymous referee for their suggestions to improve the paper.
\end{acks}

\

\small{\noindent \textbf{Disclosure of Potential Conflicts of Interest:}   The authors declare that they have no conflicts of interest.}


%
%
\bibliographystyle{spr-mp-sola}
\bibliography{bibliography.bbl}  

\begin{thebibliography}{84}
\ifx\bisbn     \undefined \def\bisbn  #1{ISBN #1}\fi
\ifx\binits    \undefined \def\binits#1{#1}\fi
\ifx\bauthor   \undefined \def\bauthor#1{#1}\fi
\ifx\batitle   \undefined \def\batitle#1{#1}\fi
\ifx\bjtitle   \undefined \def\bjtitle#1{\textit{#1}}\fi
\ifx\bvolume   \undefined \def\bvolume#1{\textbf{#1}}\fi
\ifx\byear     \undefined \def\byear#1{#1}\fi
\ifx\bissue    \undefined \def\bissue#1{#1}\fi
\ifx\bfpage    \undefined \def\bfpage#1{#1}\fi
\ifx\blpage    \undefined \def\blpage #1{#1}\fi
\ifx\burl      \undefined \def\burl#1{\textsf{#1}}\fi
\ifx\href      \undefined \def\href#1#2{\textsf{#2}}\fi
\ifx\betal     \undefined \def\betal{\textit{et al.}}\fi
\ifx\bctitle   \undefined \def\bctitle#1{#1}\fi
\ifx\beditor   \undefined \def\beditor#1{#1}\fi
\ifx\bbtitle   \undefined \def\bbtitle#1{\textit{#1}}\fi
\ifx\bedition  \undefined \def\bedition#1{#1}\fi
\ifx\bseriesno \undefined \def\bseriesno#1{\textbf{#1}}\fi
\ifx\blocation \undefined \def\blocation#1{#1}\fi
\ifx\bsertitle \undefined \def\bsertitle#1{\textit{#1}}\fi
\ifx\bsnm      \undefined \def\bsnm#1{#1}\fi
\ifx\bsuffix   \undefined \def\bsuffix#1{#1}\fi
\ifx\bparticle \undefined \def\bparticle#1{#1}\fi
\ifx\barticle  \undefined \def\barticle#1{}\fi
\ifx\binstitute  \undefined \def\binstitute#1{#1}\fi
\ifx\bpublisher  \undefined \def\bpublisher#1{#1}\fi
\ifx\doiurl    \undefined
  \def\doiurl#1{\href{http://dx.doi.org/#1}{\textsf{DOI}}}\fi
\ifx\arxivurl  \undefined
  \def\arxivurl#1{\href{http://arxiv.org/abs/#1}{\textsf{arXiv}}}\fi
\ifx\adsurl    \undefined
  \def\adsurl#1{\href{http://adsabs.harvard.edu/abs/#1}{\textsf{ADS}}}\fi
\ifx\botherref \undefined \def\botherref#1{}\fi
\ifx\url       \undefined \def\url#1{\textsf{#1}}\fi
\ifx\bchapter  \undefined \def\bchapter#1{}\fi
\ifx\bbook     \undefined \def\bbook#1{}\fi
\ifx\bcomment  \undefined \def\bcomment#1{#1}\fi
\ifx\oauthor   \undefined \def\oauthor#1{#1}\fi
\ifx\citeauthoryear \undefined\def \citeauthoryear#1{#1}\fi
\ifx\endbibitem\undefined \def\endbibitem{}\fi
\ifx\bconflocation  \undefined \def\bconflocation#1{#1} \fi

\bibitem[\protect\citeauthoryear{{Abramenko}
  \textit{et~al.}}{2002}]{2002ApJ...577..487A}
\begin{barticle}
\bauthor{\bsnm{{Abramenko}}, \binits{V.I.}},
\bauthor{\bsnm{{Yurchyshyn}}, \binits{V.B.}},
\bauthor{\bsnm{{Wang}}, \binits{H.}},
\bauthor{\bsnm{{Spirock}}, \binits{T.J.}},
\bauthor{\bsnm{{Goode}}, \binits{P.R.}}:
\byear{2002},
\batitle{{Scaling Behavior of Structure Functions of the Longitudinal Magnetic
  Field in Active Regions on the Sun}}.
\bjtitle{\apj}
\bvolume{577},
\bfpage{487}.
\doiurl{10.1086/342169}.
\adsurl{2002ApJ...577..487A}.
\end{barticle}
\endbibitem

\bibitem[\protect\citeauthoryear{Ahmed \textit{et~al.}}{2010}]{Ahmed2010}
\begin{barticle}
\bauthor{\bsnm{Ahmed}, \binits{O.W.}},
\bauthor{\bsnm{Qahwaji}, \binits{R.}},
\bauthor{\bsnm{Colak}, \binits{T.}},
\bauthor{\bsnm{Dudok€ De~Wit}, \binits{T.}},
\bauthor{\bsnm{Ipson}, \binits{S.}}:
\byear{2010},
\batitle{A new technique for the calculation and 3d visualisation
  of magnetic complexities on solar satellite images}.
\bjtitle{The Visual Computer}
\bvolume{26}(\bissue{5}),
\bfpage{385}.
\doiurl{10.1007/s00371-010-0418-1}.
\burl{https://doi.org/10.1007/s00371-010-0418-1}.
\end{barticle}
\endbibitem

\bibitem[\protect\citeauthoryear{{Ahmed} \textit{et~al.}}{2013}]{ahmed13}
\begin{barticle}
\bauthor{\bsnm{{Ahmed}}, \binits{O.W.}},
\bauthor{\bsnm{{Qahwaji}}, \binits{R.}},
\bauthor{\bsnm{{Colak}}, \binits{T.}},
\bauthor{\bsnm{{Higgins}}, \binits{P.A.}},
\bauthor{\bsnm{{Gallagher}}, \binits{P.T.}},
\bauthor{\bsnm{{Bloomfield}}, \binits{D.S.}}:
\byear{2013},
\batitle{{Solar Flare Prediction Using Advanced Feature Extraction, Machine
  Learning, and Feature Selection}}.
\bjtitle{\solphys}
\bvolume{283},
\bfpage{157}.
\doiurl{10.1007/s11207-011-9896-1}.
\adsurl{2013SoPh..283..157A}.
\end{barticle}
\endbibitem

\bibitem[\protect\citeauthoryear{{Andrews}}{2003}]{andrews03}
\begin{barticle}
\bauthor{\bsnm{{Andrews}}, \binits{M.D.}}:
\byear{2003},
\batitle{{A Search for CMEs Associated with Big Flares}}.
\bjtitle{\solphys}
\bvolume{218},
\bfpage{261}.
\doiurl{10.1023/B:SOLA.0000013039.69550.bf}.
\adsurl{2003SoPh..218..261A}.
\end{barticle}
\endbibitem

\bibitem[\protect\citeauthoryear{Barnes \textit{et~al.}}{2015}]{barnes15}
\begin{botherref}
\oauthor{\bsnm{Barnes}, \binits{D.}},
\oauthor{\bsnm{Byrne}, \binits{J.}},
\oauthor{\bsnm{Davies}, \binits{J.}},
\oauthor{\bsnm{Harrison}, \binits{R.}},
\oauthor{\bsnm{Helcats}, \binits{E.U.}}:
2015,
{HELCATS HCME\_WP2\_V02}.
\doiurl{10.6084/m9.figshare.1492351.v1}.
\url{https://figshare.com/articles/HELCATS_HCME_WP2_V02/1492351}.
\end{botherref}
\endbibitem

\bibitem[\protect\citeauthoryear{{Barnes} \textit{et~al.}}{2016}]{barnes16}
\begin{barticle}
\bauthor{\bsnm{{Barnes}}, \binits{G.}},
\bauthor{\bsnm{{Leka}}, \binits{K.D.}},
\bauthor{\bsnm{{Schrijver}}, \binits{C.J.}},
\bauthor{\bsnm{{Colak}}, \binits{T.}},
\bauthor{\bsnm{{Qahwaji}}, \binits{R.}},
\bauthor{\bsnm{{Ashamari}}, \binits{O.W.}},
\bauthor{\bsnm{{Yuan}}, \binits{Y.}},
\bauthor{\bsnm{{Zhang}}, \binits{J.}},
\bauthor{\bsnm{{McAteer}}, \binits{R.T.J.}},
\bauthor{\bsnm{{Bloomfield}}, \binits{D.S.}},
\bauthor{\bsnm{{Higgins}}, \binits{P.A.}},
\bauthor{\bsnm{{Gallagher}}, \binits{P.T.}},
\bauthor{\bsnm{{Falconer}}, \binits{D.A.}},
\bauthor{\bsnm{{Georgoulis}}, \binits{M.K.}},
\bauthor{\bsnm{{Wheatland}}, \binits{M.S.}},
\bauthor{\bsnm{{Balch}}, \binits{C.}},
\bauthor{\bsnm{{Dunn}}, \binits{T.}},
\bauthor{\bsnm{{Wagner}}, \binits{E.L.}}:
\byear{2016},
\batitle{{A Comparison of Flare Forecasting Methods. I. Results from the
  `All-Clear' Workshop}}.
\bjtitle{\apj}
\bvolume{829},
\bfpage{89}.
\doiurl{10.3847/0004-637X/829/2/89}.
\adsurl{2016ApJ...829...89B}.
\end{barticle}
\endbibitem

\bibitem[\protect\citeauthoryear{Berger and Field}{1984}]{berger_field_1984}
\begin{barticle}
\bauthor{\bsnm{Berger}, \binits{M.A.}},
\bauthor{\bsnm{Field}, \binits{G.B.}}:
\byear{1984},
\batitle{The topological properties of magnetic helicity}.
\bjtitle{Journal of Fluid Mechanics}
\bvolume{147},
\bfpage{133–148}.
\doiurl{10.1017/S0022112084002019}.
\end{barticle}
\endbibitem

\bibitem[\protect\citeauthoryear{{Berghmans}, {Foing}, and
  {Fleck}}{2002}]{berghmans02}
\begin{bchapter}
\bauthor{\bsnm{{Berghmans}}, \binits{D.}},
\bauthor{\bsnm{{Foing}}, \binits{B.H.}},
\bauthor{\bsnm{{Fleck}}, \binits{B.}}:
\byear{2002},
\bctitle{{Automated detection of CMEs in LASCO data}}.
In: \beditor{\bsnm{{Wilson}}, \binits{A.}} (ed.)
\bbtitle{From Solar Min to Max: Half a Solar Cycle with SOHO},
\bsertitle{ESA Special Publication}
\bseriesno{508},
\bfpage{437}.
\adsurl{2002ESASP.508..437B}.
\end{bchapter}
\endbibitem

\bibitem[\protect\citeauthoryear{{Bloomfield}
  \textit{et~al.}}{2012}]{bloomfield12}
\begin{barticle}
\bauthor{\bsnm{{Bloomfield}}, \binits{D.S.}},
\bauthor{\bsnm{{Higgins}}, \binits{P.A.}},
\bauthor{\bsnm{{McAteer}}, \binits{R.T.J.}},
\bauthor{\bsnm{{Gallagher}}, \binits{P.T.}}:
\byear{2012},
\batitle{{Toward Reliable Benchmarking of Solar Flare Forecasting Methods}}.
\bjtitle{\apjl}
\bvolume{747},
\bfpage{L41}.
\doiurl{10.1088/2041-8205/747/2/L41}.
\adsurl{2012ApJ...747L..41B}.
\end{barticle}
\endbibitem

\bibitem[\protect\citeauthoryear{{Bobra} \textit{et~al.}}{2014a}]{bobra14}
\begin{barticle}
\bauthor{\bsnm{{Bobra}}, \binits{M.G.}},
\bauthor{\bsnm{{Sun}}, \binits{X.}},
\bauthor{\bsnm{{Hoeksema}}, \binits{J.T.}},
\bauthor{\bsnm{{Turmon}}, \binits{M.}},
\bauthor{\bsnm{{Liu}}, \binits{Y.}},
\bauthor{\bsnm{{Hayashi}}, \binits{K.}},
\bauthor{\bsnm{{Barnes}}, \binits{G.}},
\bauthor{\bsnm{{Leka}}, \binits{K.D.}}:
\byear{2014}a,
\batitle{{The Helioseismic and Magnetic Imager (HMI) Vector Magnetic Field
  Pipeline: SHARPs - Space-Weather HMI Active Region Patches}}.
\bjtitle{\solphys}
\bvolume{289},
\bfpage{3549}.
\doiurl{10.1007/s11207-014-0529-3}.
\adsurl{2014SoPh..289.3549B}.
\end{barticle}
\endbibitem

\bibitem[\protect\citeauthoryear{{Bobra}
  \textit{et~al.}}{2014b}]{2014SoPh..289.3549B}
\begin{barticle}
\bauthor{\bsnm{{Bobra}}, \binits{M.G.}},
\bauthor{\bsnm{{Sun}}, \binits{X.}},
\bauthor{\bsnm{{Hoeksema}}, \binits{J.T.}},
\bauthor{\bsnm{{Turmon}}, \binits{M.}},
\bauthor{\bsnm{{Liu}}, \binits{Y.}},
\bauthor{\bsnm{{Hayashi}}, \binits{K.}},
\bauthor{\bsnm{{Barnes}}, \binits{G.}},
\bauthor{\bsnm{{Leka}}, \binits{K.D.}}:
\byear{2014}b,
\batitle{{The Helioseismic and Magnetic Imager (HMI) Vector Magnetic Field
  Pipeline: SHARPs - Space-Weather HMI Active Region Patches}}.
\bjtitle{\solphys}
\bvolume{289},
\bfpage{3549}.
\doiurl{10.1007/s11207-014-0529-3}.
\adsurl{2014SoPh..289.3549B}.
\end{barticle}
\endbibitem

\bibitem[\protect\citeauthoryear{{Brueckner}
  \textit{et~al.}}{1995}]{brueckner95}
\begin{barticle}
\bauthor{\bsnm{{Brueckner}}, \binits{G.E.}},
\bauthor{\bsnm{{Howard}}, \binits{R.A.}},
\bauthor{\bsnm{{Koomen}}, \binits{M.J.}},
\bauthor{\bsnm{{Korendyke}}, \binits{C.M.}},
\bauthor{\bsnm{{Michels}}, \binits{D.J.}},
\bauthor{\bsnm{{Moses}}, \binits{J.D.}},
\bauthor{\bsnm{{Socker}}, \binits{D.G.}},
\bauthor{\bsnm{{Dere}}, \binits{K.P.}},
\bauthor{\bsnm{{Lamy}}, \binits{P.L.}},
\bauthor{\bsnm{{Llebaria}}, \binits{A.}},
\bauthor{\bsnm{{Bout}}, \binits{M.V.}},
\bauthor{\bsnm{{Schwenn}}, \binits{R.}},
\bauthor{\bsnm{{Simnett}}, \binits{G.M.}},
\bauthor{\bsnm{{Bedford}}, \binits{D.K.}},
\bauthor{\bsnm{{Eyles}}, \binits{C.J.}}:
\byear{1995},
\batitle{{The Large Angle Spectroscopic Coronagraph (LASCO)}}.
\bjtitle{\solphys}
\bvolume{162},
\bfpage{357}.
\doiurl{10.1007/BF00733434}.
\adsurl{http://cdsads.u-strasbg.fr/abs/1995SoPh..162..357B}.
\end{barticle}
\endbibitem

\bibitem[\protect\citeauthoryear{{Byrne} \textit{et~al.}}{2010}]{byrne10}
\begin{barticle}
\bauthor{\bsnm{{Byrne}}, \binits{J.P.}},
\bauthor{\bsnm{{Maloney}}, \binits{S.A.}},
\bauthor{\bsnm{{McAteer}}, \binits{R.T.J.}},
\bauthor{\bsnm{{Refojo}}, \binits{J.M.}},
\bauthor{\bsnm{{Gallagher}}, \binits{P.T.}}:
\byear{2010},
\batitle{{Propagation of an Earth-directed coronal mass ejection in three
  dimensions}}.
\bjtitle{Nature Communications}
\bvolume{1},
\bfpage{74}.
\doiurl{10.1038/ncomms1077}.
\adsurl{2010NatCo...1E..74B}.
\end{barticle}
\endbibitem

\bibitem[\protect\citeauthoryear{{Byrne} \textit{et~al.}}{2012}]{byrne12}
\begin{barticle}
\bauthor{\bsnm{{Byrne}}, \binits{J.P.}},
\bauthor{\bsnm{{Morgan}}, \binits{H.}},
\bauthor{\bsnm{{Habbal}}, \binits{S.R.}},
\bauthor{\bsnm{{Gallagher}}, \binits{P.T.}}:
\byear{2012},
\batitle{{Automatic Detection and Tracking of Coronal Mass Ejections. II.
  Multiscale Filtering of Coronagraph Images}}.
\bjtitle{\apj}
\bvolume{752},
\bfpage{145}.
\doiurl{10.1088/0004-637X/752/2/145}.
\adsurl{2012ApJ...752..145B}.
\end{barticle}
\endbibitem

\bibitem[\protect\citeauthoryear{Campi and Benvenuto}{2017}]{CristinaUNIGE}
\begin{botherref}
\oauthor{\bsnm{Campi}, \binits{C.}},
\oauthor{\bsnm{Benvenuto}, \binits{F.}}:
2017,
Feature selection for flarecast.
\textit{private communication}.
\end{botherref}
\endbibitem

\bibitem[\protect\citeauthoryear{Conlon \textit{et~al.}}{2008}]{Conlon2008}
\begin{barticle}
\bauthor{\bsnm{Conlon}, \binits{P.A.}},
\bauthor{\bsnm{Gallagher}, \binits{P.T.}},
\bauthor{\bsnm{McAteer}, \binits{R.T.J.}},
\bauthor{\bsnm{Ireland}, \binits{J.}},
\bauthor{\bsnm{Young}, \binits{C.A.}},
\bauthor{\bsnm{Kestener}, \binits{P.}},
\bauthor{\bsnm{Hewett}, \binits{R.J.}},
\bauthor{\bsnm{Maguire}, \binits{K.}}:
\byear{2008},
\batitle{Multifractal properties of evolving active regions}.
\bjtitle{\solphys}
\bvolume{248}(\bissue{2}),
\bfpage{297}.
\doiurl{10.1007/s11207-007-9074-7}.
\burl{https://doi.org/10.1007/s11207-007-9074-7}.
\end{barticle}
\endbibitem

\bibitem[\protect\citeauthoryear{{Davies} \textit{et~al.}}{2012}]{davies12}
\begin{barticle}
\bauthor{\bsnm{{Davies}}, \binits{J.A.}},
\bauthor{\bsnm{{Harrison}}, \binits{R.A.}},
\bauthor{\bsnm{{Perry}}, \binits{C.H.}},
\bauthor{\bsnm{{M{\"o}stl}}, \binits{C.}},
\bauthor{\bsnm{{Lugaz}}, \binits{N.}},
\bauthor{\bsnm{{Rollett}}, \binits{T.}},
\bauthor{\bsnm{{Davis}}, \binits{C.J.}},
\bauthor{\bsnm{{Crothers}}, \binits{S.R.}},
\bauthor{\bsnm{{Temmer}}, \binits{M.}},
\bauthor{\bsnm{{Eyles}}, \binits{C.J.}},
\bauthor{\bsnm{{Savani}}, \binits{N.P.}}:
\byear{2012},
\batitle{{A Self-similar Expansion Model for Use in Solar Wind Transient
  Propagation Studies}}.
\bjtitle{\apj}
\bvolume{750},
\bfpage{23}.
\doiurl{10.1088/0004-637X/750/1/23}.
\adsurl{2012ApJ...750...23D}.
\end{barticle}
\endbibitem

\bibitem[\protect\citeauthoryear{{Davies} \textit{et~al.}}{2013}]{davies13}
\begin{barticle}
\bauthor{\bsnm{{Davies}}, \binits{J.A.}},
\bauthor{\bsnm{{Perry}}, \binits{C.H.}},
\bauthor{\bsnm{{Trines}}, \binits{R.M.G.M.}},
\bauthor{\bsnm{{Harrison}}, \binits{R.A.}},
\bauthor{\bsnm{{Lugaz}}, \binits{N.}},
\bauthor{\bsnm{{M{\"o}stl}}, \binits{C.}},
\bauthor{\bsnm{{Liu}}, \binits{Y.D.}},
\bauthor{\bsnm{{Steed}}, \binits{K.}}:
\byear{2013},
\batitle{{Establishing a Stereoscopic Technique for Determining the Kinematic
  Properties of Solar Wind Transients based on a Generalized Self-similarly
  Expanding Circular Geometry}}.
\bjtitle{\apj}
\bvolume{777},
\bfpage{167}.
\doiurl{10.1088/0004-637X/777/2/167}.
\adsurl{2013ApJ...777..167D}.
\end{barticle}
\endbibitem

\bibitem[\protect\citeauthoryear{Deng \textit{et~al.}}{2006}]{Deng2006}
\begin{barticle}
\bauthor{\bsnm{Deng}, \binits{N.}},
\bauthor{\bsnm{Xu}, \binits{Y.}},
\bauthor{\bsnm{Yang}, \binits{G.}},
\bauthor{\bsnm{Cao}, \binits{W.}},
\bauthor{\bsnm{Liu}, \binits{C.}},
\bauthor{\bsnm{Rimmele}, \binits{T.R.}},
\bauthor{\bsnm{Wang}, \binits{H.}},
\bauthor{\bsnm{Denker}, \binits{C.}}:
\byear{2006},
\batitle{Multiwavelength study of flow fields in flaring super active region
  noaa 10486}.
\bjtitle{\apj}
\bvolume{644}(\bissue{2}),
\bfpage{1278}.
\burl{http://stacks.iop.org/0004-637X/644/i=2/a=1278}.
\end{barticle}
\endbibitem

\bibitem[\protect\citeauthoryear{{Drake}}{1971}]{drake71}
\begin{barticle}
\bauthor{\bsnm{{Drake}}, \binits{J.F.}}:
\byear{1971},
\batitle{{Characteristics of Soft Solar X-Ray Bursts}}.
\bjtitle{\solphys}
\bvolume{16},
\bfpage{152}.
\doiurl{10.1007/BF00154510}.
\adsurl{1971SoPh...16..152D}.
\end{barticle}
\endbibitem

\bibitem[\protect\citeauthoryear{Eastwood \textit{et~al.}}{2017}]{eastwood17}
\begin{barticle}
\bauthor{\bsnm{Eastwood}, \binits{J.P.}},
\bauthor{\bsnm{Biffis}, \binits{E.}},
\bauthor{\bsnm{Hapgood}, \binits{M.A.}},
\bauthor{\bsnm{Green}, \binits{L.}},
\bauthor{\bsnm{Bisi}, \binits{M.M.}},
\bauthor{\bsnm{Bentley}, \binits{R.D.}},
\bauthor{\bsnm{Wicks}, \binits{R.}},
\bauthor{\bsnm{McKinnell}, \binits{L.-A.}},
\bauthor{\bsnm{Gibbs}, \binits{M.}},
\bauthor{\bsnm{Burnett}, \binits{C.}}:
\byear{2017},
\batitle{The economic impact of space weather: Where do we stand?}
\bjtitle{Risk Analysis}
\bvolume{37}(\bissue{2}),
\bfpage{206}.
\doiurl{10.1111/risa.12765}.
\burl{http://dx.doi.org/10.1111/risa.12765}.
\end{barticle}
\endbibitem

\bibitem[\protect\citeauthoryear{{Eyles} \textit{et~al.}}{2009}]{eyles09}
\begin{barticle}
\bauthor{\bsnm{{Eyles}}, \binits{C.J.}},
\bauthor{\bsnm{{Harrison}}, \binits{R.A.}},
\bauthor{\bsnm{{Davis}}, \binits{C.J.}},
\bauthor{\bsnm{{Waltham}}, \binits{N.R.}},
\bauthor{\bsnm{{Shaughnessy}}, \binits{B.M.}},
\bauthor{\bsnm{{Mapson-Menard}}, \binits{H.C.A.}},
\bauthor{\bsnm{{Bewsher}}, \binits{D.}},
\bauthor{\bsnm{{Crothers}}, \binits{S.R.}},
\bauthor{\bsnm{{Davies}}, \binits{J.A.}},
\bauthor{\bsnm{{Simnett}}, \binits{G.M.}},
\bauthor{\bsnm{{Howard}}, \binits{R.A.}},
\bauthor{\bsnm{{Moses}}, \binits{J.D.}},
\bauthor{\bsnm{{Newmark}}, \binits{J.S.}},
\bauthor{\bsnm{{Socker}}, \binits{D.G.}},
\bauthor{\bsnm{{Halain}}, \binits{J.-P.}},
\bauthor{\bsnm{{Defise}}, \binits{J.-M.}},
\bauthor{\bsnm{{Mazy}}, \binits{E.}},
\bauthor{\bsnm{{Rochus}}, \binits{P.}}:
\byear{2009},
\batitle{{The Heliospheric Imagers Onboard the STEREO Mission}}.
\bjtitle{\solphys}
\bvolume{254},
\bfpage{387}.
\doiurl{10.1007/s11207-008-9299-0}.
\adsurl{2009SoPh..254..387E}.
\end{barticle}
\endbibitem

\bibitem[\protect\citeauthoryear{{Falconer}, {Moore}, and
  {Gary}}{2008}]{falconer08}
\begin{barticle}
\bauthor{\bsnm{{Falconer}}, \binits{D.A.}},
\bauthor{\bsnm{{Moore}}, \binits{R.L.}},
\bauthor{\bsnm{{Gary}}, \binits{G.A.}}:
\byear{2008},
\batitle{{Magnetogram Measures of Total Nonpotentiality for Prediction of Solar
  Coronal Mass Ejections from Active Regions of Any Degree of Magnetic
  Complexity}}.
\bjtitle{\apj}
\bvolume{689},
\bfpage{1433}.
\doiurl{10.1086/591045}.
\adsurl{2008ApJ...689.1433F}.
\end{barticle}
\endbibitem

\bibitem[\protect\citeauthoryear{{Falconer}
  \textit{et~al.}}{2012}]{2012ApJ...757...32F}
\begin{barticle}
\bauthor{\bsnm{{Falconer}}, \binits{D.A.}},
\bauthor{\bsnm{{Moore}}, \binits{R.L.}},
\bauthor{\bsnm{{Barghouty}}, \binits{A.F.}},
\bauthor{\bsnm{{Khazanov}}, \binits{I.}}:
\byear{2012},
\batitle{{Prior Flaring as a Complement to Free Magnetic Energy for Forecasting
  Solar Eruptions}}.
\bjtitle{\apj}
\bvolume{757},
\bfpage{32}.
\doiurl{10.1088/0004-637X/757/1/32}.
\adsurl{2012ApJ...757...32F}.
\end{barticle}
\endbibitem

\bibitem[\protect\citeauthoryear{Georgoulis}{2013}]{Georgoulis_2013}
\begin{barticle}
\bauthor{\bsnm{Georgoulis}, \binits{M.}}:
\byear{2013},
\batitle{Toward an efficient prediction of solar flares: Which parameters, and
  how?}
\bjtitle{Entropy}
\bvolume{15}(\bissue{11}),
\bfpage{5022}.
\doiurl{10.3390/e15115022}.
\burl{http://dx.doi.org/10.3390/e15115022}.
\end{barticle}
\endbibitem

\bibitem[\protect\citeauthoryear{Georgoulis}{2010}]{georgoulis2010}
\begin{botherref}
\oauthor{\bsnm{Georgoulis}, \binits{M.K.}}:
2010,
Pre-eruption magnetic configurations in the active-region solar photosphere.
\textbf{6}(S273),
495.
\doiurl{10.1017/S1743921311015870}.
\end{botherref}
\endbibitem

\bibitem[\protect\citeauthoryear{{Georgoulis} and {Rust}}{2007}]{georgoulis07}
\begin{barticle}
\bauthor{\bsnm{{Georgoulis}}, \binits{M.K.}},
\bauthor{\bsnm{{Rust}}, \binits{D.M.}}:
\byear{2007},
\batitle{{Quantitative Forecasting of Major Solar Flares}}.
\bjtitle{\apjl}
\bvolume{661},
\bfpage{L109}.
\doiurl{10.1086/518718}.
\adsurl{2007ApJ...661L.109G}.
\end{barticle}
\endbibitem

\bibitem[\protect\citeauthoryear{Georgoulis, Titov, and
  Mikić}{2012}]{Georgoulis2012}
\begin{barticle}
\bauthor{\bsnm{Georgoulis}, \binits{M.K.}},
\bauthor{\bsnm{Titov}, \binits{V.S.}},
\bauthor{\bsnm{Mikić}, \binits{Z.}}:
\byear{2012},
\batitle{Non-neutralized electric current patterns in solar active regions:
  Origin of the shear-generating lorentz force}.
\bjtitle{\apj}
\bvolume{761}(\bissue{1}),
\bfpage{61}.
\burl{http://stacks.iop.org/0004-637X/761/i=1/a=61}.
\end{barticle}
\endbibitem

\bibitem[\protect\citeauthoryear{{Gopalswamy}
  \textit{et~al.}}{2009}]{gopalswamy09}
\begin{barticle}
\bauthor{\bsnm{{Gopalswamy}}, \binits{N.}},
\bauthor{\bsnm{{Yashiro}}, \binits{S.}},
\bauthor{\bsnm{{Michalek}}, \binits{G.}},
\bauthor{\bsnm{{Stenborg}}, \binits{G.}},
\bauthor{\bsnm{{Vourlidas}}, \binits{A.}},
\bauthor{\bsnm{{Freeland}}, \binits{S.}},
\bauthor{\bsnm{{Howard}}, \binits{R.}}:
\byear{2009},
\batitle{{The SOHO/LASCO CME Catalog}}.
\bjtitle{Earth Moon and Planets}
\bvolume{104},
\bfpage{295}.
\doiurl{10.1007/s11038-008-9282-7}.
\adsurl{2009EM\%26P..104..295G}.
\end{barticle}
\endbibitem

\bibitem[\protect\citeauthoryear{{Guerra}
  \textit{et~al.}}{2015}]{2015SoPh..290..335G}
\begin{barticle}
\bauthor{\bsnm{{Guerra}}, \binits{J.A.}},
\bauthor{\bsnm{{Pulkkinen}}, \binits{A.}},
\bauthor{\bsnm{{Uritsky}}, \binits{V.M.}},
\bauthor{\bsnm{{Yashiro}}, \binits{S.}}:
\byear{2015},
\batitle{{Spatio-Temporal Scaling of Turbulent Photospheric Line-of-Sight
  Magnetic Field in Active Region NOAA 11158}}.
\bjtitle{\solphys}
\bvolume{290},
\bfpage{335}.
\doiurl{10.1007/s11207-014-0636-1}.
\adsurl{2015SoPh..290..335G}.
\end{barticle}
\endbibitem

\bibitem[\protect\citeauthoryear{{Guo}, {Zhang}, and {Chumak}}{2007}]{guo07}
\begin{barticle}
\bauthor{\bsnm{{Guo}}, \binits{J.}},
\bauthor{\bsnm{{Zhang}}, \binits{H.Q.}},
\bauthor{\bsnm{{Chumak}}, \binits{O.V.}}:
\byear{2007},
\batitle{{Magnetic properties of flare-CME productive active regions and CME
  speed}}.
\bjtitle{\aap}
\bvolume{462},
\bfpage{1121}.
\doiurl{10.1051/0004-6361:20065888}.
\adsurl{2007A\%26A...462.1121G}.
\end{barticle}
\endbibitem

\bibitem[\protect\citeauthoryear{{Harra} \textit{et~al.}}{2016}]{harra16}
\begin{barticle}
\bauthor{\bsnm{{Harra}}, \binits{L.K.}},
\bauthor{\bsnm{{Schrijver}}, \binits{C.J.}},
\bauthor{\bsnm{{Janvier}}, \binits{M.}},
\bauthor{\bsnm{{Toriumi}}, \binits{S.}},
\bauthor{\bsnm{{Hudson}}, \binits{H.}},
\bauthor{\bsnm{{Matthews}}, \binits{S.}},
\bauthor{\bsnm{{Woods}}, \binits{M.M.}},
\bauthor{\bsnm{{Hara}}, \binits{H.}},
\bauthor{\bsnm{{Guedel}}, \binits{M.}},
\bauthor{\bsnm{{Kowalski}}, \binits{A.}},
\bauthor{\bsnm{{Osten}}, \binits{R.}},
\bauthor{\bsnm{{Kusano}}, \binits{K.}},
\bauthor{\bsnm{{Lueftinger}}, \binits{T.}}:
\byear{2016},
\batitle{{The Characteristics of Solar X-Class Flares and CMEs: A Paradigm for
  Stellar Superflares and Eruptions?}}
\bjtitle{\solphys}
\bvolume{291},
\bfpage{1761}.
\doiurl{10.1007/s11207-016-0923-0}.
\adsurl{2016SoPh..291.1761H}.
\end{barticle}
\endbibitem

\bibitem[\protect\citeauthoryear{{Harrison}}{1995}]{harrison95}
\begin{barticle}
\bauthor{\bsnm{{Harrison}}, \binits{R.A.}}:
\byear{1995},
\batitle{{The nature of solar flares associated with coronal mass ejection.}}
\bjtitle{\aap}
\bvolume{304},
\bfpage{585}.
\adsurl{1995A\%26A...304..585H}.
\end{barticle}
\endbibitem

\bibitem[\protect\citeauthoryear{Harrison \textit{et~al.}}{2017}]{harrison17}
\begin{barticle}
\bauthor{\bsnm{Harrison}, \binits{R.A.}},
\bauthor{\bsnm{Davies}, \binits{J.A.}},
\bauthor{\bsnm{Biesecker}, \binits{D.}},
\bauthor{\bsnm{Gibbs}, \binits{M.}}:
\byear{2017},
\batitle{The application of heliospheric imaging to space weather operations:
  Lessons learned from published studies}.
\bjtitle{Space Weather}
\bvolume{15}(\bissue{8}),
\bfpage{985}.
\bcomment{2017SW001633}.
\doiurl{10.1002/2017SW001633}.
\burl{http://dx.doi.org/10.1002/2017SW001633}.
\end{barticle}
\endbibitem

\bibitem[\protect\citeauthoryear{Haynes and Parnell}{2007}]{Haynes2007}
\begin{barticle}
\bauthor{\bsnm{Haynes}, \binits{A.L.}},
\bauthor{\bsnm{Parnell}, \binits{C.E.}}:
\byear{2007},
\batitle{A trilinear method for finding null points in a three-dimensional
  vector space}.
\bjtitle{Physics of Plasmas}
\bvolume{14}(\bissue{8}),
\bfpage{082107}.
\doiurl{10.1063/1.2756751}.
\burl{http://dx.doi.org/10.1063/1.2756751}.
\end{barticle}
\endbibitem

\bibitem[\protect\citeauthoryear{{Hewett}
  \textit{et~al.}}{2008}]{2008SoPh..248..311H}
\begin{barticle}
\bauthor{\bsnm{{Hewett}}, \binits{R.J.}},
\bauthor{\bsnm{{Gallagher}}, \binits{P.T.}},
\bauthor{\bsnm{{McAteer}}, \binits{R.T.J.}},
\bauthor{\bsnm{{Young}}, \binits{C.A.}},
\bauthor{\bsnm{{Ireland}}, \binits{J.}},
\bauthor{\bsnm{{Conlon}}, \binits{P.A.}},
\bauthor{\bsnm{{Maguire}}, \binits{K.}}:
\byear{2008},
\batitle{{Multiscale Analysis of Active Region Evolution}}.
\bjtitle{\solphys}
\bvolume{248},
\bfpage{311}.
\doiurl{10.1007/s11207-007-9028-0}.
\adsurl{2008SoPh..248..311H}.
\end{barticle}
\endbibitem

\bibitem[\protect\citeauthoryear{{Higgins}}{2012}]{higgins12}
\begin{botherref}
\oauthor{\bsnm{{Higgins}}, \binits{P.A.}}:
2012,
{Sunspot group evolution and the global magnetic field of the Sun}.
\textit{Thesis, Trinity College Dublin}.
\end{botherref}
\endbibitem

\bibitem[\protect\citeauthoryear{{Higgins} \textit{et~al.}}{2011}]{higgins11}
\begin{barticle}
\bauthor{\bsnm{{Higgins}}, \binits{P.A.}},
\bauthor{\bsnm{{Gallagher}}, \binits{P.T.}},
\bauthor{\bsnm{{McAteer}}, \binits{R.T.J.}},
\bauthor{\bsnm{{Bloomfield}}, \binits{D.S.}}:
\byear{2011},
\batitle{{Solar magnetic feature detection and tracking for space weather
  monitoring}}.
\bjtitle{Advances in Space Research}
\bvolume{47},
\bfpage{2105}.
\doiurl{10.1016/j.asr.2010.06.024}.
\adsurl{2011AdSpR..47.2105H}.
\end{barticle}
\endbibitem

\bibitem[\protect\citeauthoryear{{Jing} \textit{et~al.}}{2004}]{jing04}
\begin{barticle}
\bauthor{\bsnm{{Jing}}, \binits{J.}},
\bauthor{\bsnm{{Yurchyshyn}}, \binits{V.B.}},
\bauthor{\bsnm{{Yang}}, \binits{G.}},
\bauthor{\bsnm{{Xu}}, \binits{Y.}},
\bauthor{\bsnm{{Wang}}, \binits{H.}}:
\byear{2004},
\batitle{{On the Relation between Filament Eruptions, Flares, and Coronal Mass
  Ejections}}.
\bjtitle{\apj}
\bvolume{614},
\bfpage{1054}.
\doiurl{10.1086/423781}.
\adsurl{2004ApJ...614.1054J}.
\end{barticle}
\endbibitem

\bibitem[\protect\citeauthoryear{{Kaiser} \textit{et~al.}}{2008}]{kaiser08}
\begin{barticle}
\bauthor{\bsnm{{Kaiser}}, \binits{M.L.}},
\bauthor{\bsnm{{Kucera}}, \binits{T.A.}},
\bauthor{\bsnm{{Davila}}, \binits{J.M.}},
\bauthor{\bsnm{{St.~Cyr}}, \binits{O.C.}},
\bauthor{\bsnm{{Guhathakurta}}, \binits{M.}},
\bauthor{\bsnm{{Christian}}, \binits{E.}}:
\byear{2008},
\batitle{{The STEREO Mission: An Introduction}}.
\bjtitle{\ssr}
\bvolume{136},
\bfpage{5}.
\doiurl{10.1007/s11214-007-9277-0}.
\adsurl{2008SSRv..136....5K}.
\end{barticle}
\endbibitem

\bibitem[\protect\citeauthoryear{Kors\'os, Baranyi, and
  Ludm√°ny}{2014}]{Korsos2014}
\begin{barticle}
\bauthor{\bsnm{Kors\'os}, \binits{M.B.}},
\bauthor{\bsnm{Baranyi}, \binits{T.}},
\bauthor{\bsnm{Ludm√°ny}, \binits{A.}}:
\byear{2014},
\batitle{Pre-flare dynamics of sunspot groups}.
\bjtitle{\apj}
\bvolume{789}(\bissue{2}),
\bfpage{107}.
\burl{http://stacks.iop.org/0004-637X/789/i=2/a=107}.
\end{barticle}
\endbibitem

\bibitem[\protect\citeauthoryear{{K{\"u}nzel}}{1965}]{kunzel65}
\begin{barticle}
\bauthor{\bsnm{{K{\"u}nzel}}, \binits{H.}}:
\byear{1965},
\batitle{{Zur Klassifikation von Sonnenfleckengruppen}}.
\bjtitle{Astronomische Nachrichten}
\bvolume{288},
\bfpage{177}.
\adsurl{1965AN....288..177K}.
\end{barticle}
\endbibitem

\bibitem[\protect\citeauthoryear{Kusano \textit{et~al.}}{2002}]{Kusano2002}
\begin{barticle}
\bauthor{\bsnm{Kusano}, \binits{K.}},
\bauthor{\bsnm{Maeshiro}, \binits{T.}},
\bauthor{\bsnm{Yokoyama}, \binits{T.}},
\bauthor{\bsnm{Sakurai}, \binits{T.}}:
\byear{2002},
\batitle{Measurement of magnetic helicity injection and free energy loading
  into the solar corona}.
\bjtitle{\apj}
\bvolume{577}(\bissue{1}),
\bfpage{501}.
\burl{http://stacks.iop.org/0004-637X/577/i=1/a=501}.
\end{barticle}
\endbibitem

\bibitem[\protect\citeauthoryear{{Lara}}{2008}]{lara08}
\begin{barticle}
\bauthor{\bsnm{{Lara}}, \binits{A.}}:
\byear{2008},
\batitle{{The Source Region of Coronal Mass Ejections}}.
\bjtitle{\apj}
\bvolume{688},
\bfpage{647}.
\doiurl{10.1086/591725}.
\adsurl{2008ApJ...688..647L}.
\end{barticle}
\endbibitem

\bibitem[\protect\citeauthoryear{{Lee}, {Moon}, and {Lee}}{2015}]{lee15}
\begin{barticle}
\bauthor{\bsnm{{Lee}}, \binits{K.}},
\bauthor{\bsnm{{Moon}}, \binits{Y.-J.}},
\bauthor{\bsnm{{Lee}}, \binits{J.-Y.}}:
\byear{2015},
\batitle{{Forecast of a Daily Halo CME Occurrence Probability Depending on
  Class and Area Change of the Associated Sunspot}}.
\bjtitle{\solphys}
\bvolume{290},
\bfpage{1661}.
\doiurl{10.1007/s11207-015-0715-y}.
\adsurl{2015SoPh..290.1661L}.
\end{barticle}
\endbibitem

\bibitem[\protect\citeauthoryear{Liu}{2008}]{Liu2008}
\begin{barticle}
\bauthor{\bsnm{Liu}, \binits{Y.}}:
\byear{2008},
\batitle{Magnetic field overlying solar eruption regions and kink and torus
  instabilities}.
\bjtitle{\apjl}
\bvolume{679}(\bissue{2}),
\bfpage{L151}.
\burl{http://stacks.iop.org/1538-4357/679/i=2/a=L151}.
\end{barticle}
\endbibitem

\bibitem[\protect\citeauthoryear{{Magdaleni{\'c}}
  \textit{et~al.}}{2014}]{magdalenic14}
\begin{barticle}
\bauthor{\bsnm{{Magdaleni{\'c}}}, \binits{J.}},
\bauthor{\bsnm{{Marqu{\'e}}}, \binits{C.}},
\bauthor{\bsnm{{Krupar}}, \binits{V.}},
\bauthor{\bsnm{{Mierla}}, \binits{M.}},
\bauthor{\bsnm{{Zhukov}}, \binits{A.N.}},
\bauthor{\bsnm{{Rodriguez}}, \binits{L.}},
\bauthor{\bsnm{{Maksimovi{\'c}}}, \binits{M.}},
\bauthor{\bsnm{{Cecconi}}, \binits{B.}}:
\byear{2014},
\batitle{{Tracking the CME-driven Shock Wave on 2012 March 5 and Radio
  Triangulation of Associated Radio Emission}}.
\bjtitle{\apj}
\bvolume{791},
\bfpage{115}.
\doiurl{10.1088/0004-637X/791/2/115}.
\adsurl{2014ApJ...791..115M}.
\end{barticle}
\endbibitem

\bibitem[\protect\citeauthoryear{{Mason} and
  {Hoeksema}}{2010}]{2010ApJ...723..634M}
\begin{barticle}
\bauthor{\bsnm{{Mason}}, \binits{J.P.}},
\bauthor{\bsnm{{Hoeksema}}, \binits{J.T.}}:
\byear{2010},
\batitle{{Testing Automated Solar Flare Forecasting with 13 Years of Michelson
  Doppler Imager Magnetograms}}.
\bjtitle{\apj}
\bvolume{723},
\bfpage{634}.
\doiurl{10.1088/0004-637X/723/1/634}.
\adsurl{2010ApJ...723..634M}.
\end{barticle}
\endbibitem

\bibitem[\protect\citeauthoryear{{McCauley} \textit{et~al.}}{2015}]{mccauley15}
\begin{barticle}
\bauthor{\bsnm{{McCauley}}, \binits{P.I.}},
\bauthor{\bsnm{{Su}}, \binits{Y.N.}},
\bauthor{\bsnm{{Schanche}}, \binits{N.}},
\bauthor{\bsnm{{Evans}}, \binits{K.E.}},
\bauthor{\bsnm{{Su}}, \binits{C.}},
\bauthor{\bsnm{{McKillop}}, \binits{S.}},
\bauthor{\bsnm{{Reeves}}, \binits{K.K.}}:
\byear{2015},
\batitle{{Prominence and Filament Eruptions Observed by the Solar Dynamics
  Observatory: Statistical Properties, Kinematics, and Online Catalog}}.
\bjtitle{\solphys}
\bvolume{290},
\bfpage{1703}.
\doiurl{10.1007/s11207-015-0699-7}.
\adsurl{2015SoPh..290.1703M}.
\end{barticle}
\endbibitem

\bibitem[\protect\citeauthoryear{{McIntosh}}{1990}]{mcintosh90}
\begin{barticle}
\bauthor{\bsnm{{McIntosh}}, \binits{P.S.}}:
\byear{1990},
\batitle{{The classification of sunspot groups}}.
\bjtitle{\solphys}
\bvolume{125},
\bfpage{251}.
\doiurl{10.1007/BF00158405}.
\adsurl{1990SoPh..125..251M}.
\end{barticle}
\endbibitem

\bibitem[\protect\citeauthoryear{Moestl \textit{et~al.}}{2017}]{mostl17}
\begin{barticle}
\bauthor{\bsnm{Moestl}, \binits{C.}},
\bauthor{\bsnm{Isavnin}, \binits{A.}},
\bauthor{\bsnm{Boakes}, \binits{P.D.}},
\bauthor{\bsnm{Kilpua}, \binits{E.K.J.}},
\bauthor{\bsnm{Davies}, \binits{J.A.}},
\bauthor{\bsnm{Harrison}, \binits{R.A.}},
\bauthor{\bsnm{Barnes}, \binits{D.}},
\bauthor{\bsnm{Krupar}, \binits{V.}},
\bauthor{\bsnm{Eastwood}, \binits{J.P.}},
\bauthor{\bsnm{Good}, \binits{S.W.}},
\bauthor{\bsnm{Forsyth}, \binits{R.J.}},
\bauthor{\bsnm{Bothmer}, \binits{V.}},
\bauthor{\bsnm{Reiss}, \binits{M.A.}},
\bauthor{\bsnm{Amerstorfer}, \binits{T.}},
\bauthor{\bsnm{Winslow}, \binits{R.M.}},
\bauthor{\bsnm{Anderson}, \binits{B.J.}},
\bauthor{\bsnm{Philpott}, \binits{L.C.}},
\bauthor{\bsnm{Rodriguez}, \binits{L.}},
\bauthor{\bsnm{Rouillard}, \binits{A.P.}},
\bauthor{\bsnm{Gallagher}, \binits{P.}},
\bauthor{\bsnm{Nieves-Chinchilla}, \binits{T.}},
\bauthor{\bsnm{Zhang}, \binits{T.L.}}:
\byear{2017},
\batitle{Modeling observations of solar coronal mass ejections with
  heliospheric imagers verified with the heliophysics system observatory}.
\bjtitle{Space Weather}
\bvolume{15}(\bissue{7}),
\bfpage{955}.
\bcomment{2017SW001614}.
\doiurl{10.1002/2017SW001614}.
\burl{http://dx.doi.org/10.1002/2017SW001614}.
\end{barticle}
\endbibitem

\bibitem[\protect\citeauthoryear{{Moon} \textit{et~al.}}{2002}]{moon02}
\begin{barticle}
\bauthor{\bsnm{{Moon}}, \binits{Y.-J.}},
\bauthor{\bsnm{{Choe}}, \binits{G.S.}},
\bauthor{\bsnm{{Wang}}, \binits{H.}},
\bauthor{\bsnm{{Park}}, \binits{Y.D.}},
\bauthor{\bsnm{{Gopalswamy}}, \binits{N.}},
\bauthor{\bsnm{{Yang}}, \binits{G.}},
\bauthor{\bsnm{{Yashiro}}, \binits{S.}}:
\byear{2002},
\batitle{{A Statistical Study of Two Classes of Coronal Mass Ejections}}.
\bjtitle{\apj}
\bvolume{581},
\bfpage{694}.
\doiurl{10.1086/344088}.
\adsurl{2002ApJ...581..694M}.
\end{barticle}
\endbibitem

\bibitem[\protect\citeauthoryear{{Morgan}, {Byrne}, and
  {Habbal}}{2012}]{morgan12}
\begin{barticle}
\bauthor{\bsnm{{Morgan}}, \binits{H.}},
\bauthor{\bsnm{{Byrne}}, \binits{J.P.}},
\bauthor{\bsnm{{Habbal}}, \binits{S.R.}}:
\byear{2012},
\batitle{{Automatically Detecting and Tracking Coronal Mass Ejections. I.
  Separation of Dynamic and Quiescent Components in Coronagraph Images}}.
\bjtitle{\apj}
\bvolume{752},
\bfpage{144}.
\doiurl{10.1088/0004-637X/752/2/144}.
\adsurl{2012ApJ...752..144M}.
\end{barticle}
\endbibitem

\bibitem[\protect\citeauthoryear{{Murray} \textit{et~al.}}{2017a}]{murray17}
\begin{barticle}
\bauthor{\bsnm{{Murray}}, \binits{S.A.}},
\bauthor{\bsnm{{Bingham}}, \binits{S.}},
\bauthor{\bsnm{{Sharpe}}, \binits{M.}},
\bauthor{\bsnm{{Jackson}}, \binits{D.R.}}:
\byear{2017}a,
\batitle{{Flare forecasting at the Met Office Space Weather Operations
  Centre}}.
\bjtitle{Space Weather}
\bvolume{15},
\bfpage{577}.
\doiurl{10.1002/2016SW001579}.
\adsurl{2017SpWea..15..577M}.
\end{barticle}
\endbibitem

\bibitem[\protect\citeauthoryear{{Murray} \textit{et~al.}}{2017b}]{murray17a}
\begin{botherref}
\oauthor{\bsnm{{Murray}}, \binits{S.A.}},
\oauthor{\bsnm{{Zucca}}, \binits{P.}},
\oauthor{\bsnm{{Carley}}, \binits{E.}},
\oauthor{\bsnm{{Gallagher}}, \binits{P.}}:
2017b,
{HELCATS LOWCAT}.
\textit{figshare}.
\doiurl{10.6084/m9.figshare.4970222.v2}.
\end{botherref}
\endbibitem

\bibitem[\protect\citeauthoryear{{Pant} \textit{et~al.}}{2016}]{pant16}
\begin{barticle}
\bauthor{\bsnm{{Pant}}, \binits{V.}},
\bauthor{\bsnm{{Willems}}, \binits{S.}},
\bauthor{\bsnm{{Rodriguez}}, \binits{L.}},
\bauthor{\bsnm{{Mierla}}, \binits{M.}},
\bauthor{\bsnm{{Banerjee}}, \binits{D.}},
\bauthor{\bsnm{{Davies}}, \binits{J.A.}}:
\byear{2016},
\batitle{{Automated Detection of Coronal Mass Ejections in STEREO Heliospheric
  Imager Data}}.
\bjtitle{\apj}
\bvolume{833},
\bfpage{80}.
\doiurl{10.3847/1538-4357/833/1/80}.
\adsurl{2016ApJ...833...80P}.
\end{barticle}
\endbibitem

\bibitem[\protect\citeauthoryear{Park, Chae, and Wang}{2010}]{Park2010}
\begin{barticle}
\bauthor{\bsnm{Park}, \binits{S.-H.}},
\bauthor{\bsnm{Chae}, \binits{J.}},
\bauthor{\bsnm{Wang}, \binits{H.}}:
\byear{2010},
\batitle{Productivity of solar flares and magnetic helicity injection in active
  regions}.
\bjtitle{\apj}
\bvolume{718}(\bissue{1}),
\bfpage{43}.
\burl{http://stacks.iop.org/0004-637X/718/i=1/a=43}.
\end{barticle}
\endbibitem

\bibitem[\protect\citeauthoryear{Park \textit{et~al.}}{2012}]{Park2012}
\begin{barticle}
\bauthor{\bsnm{Park}, \binits{S.-H.}},
\bauthor{\bsnm{Cho}, \binits{K.-S.}},
\bauthor{\bsnm{Bong}, \binits{S.-C.}},
\bauthor{\bsnm{Kumar}, \binits{P.}},
\bauthor{\bsnm{Chae}, \binits{J.}},
\bauthor{\bsnm{Liu}, \binits{R.}},
\bauthor{\bsnm{Wang}, \binits{H.}}:
\byear{2012},
\batitle{The occurrence and speed of cmes related to two characteristic
  evolution patterns of helicity injection in their solar source regions}.
\bjtitle{\apj}
\bvolume{750}(\bissue{1}),
\bfpage{48}.
\burl{http://stacks.iop.org/0004-637X/750/i=1/a=48}.
\end{barticle}
\endbibitem

\bibitem[\protect\citeauthoryear{{Park} \textit{et~al.}}{2012}]{park12}
\begin{barticle}
\bauthor{\bsnm{{Park}}, \binits{S.-H.}},
\bauthor{\bsnm{{Cho}}, \binits{K.-S.}},
\bauthor{\bsnm{{Bong}}, \binits{S.-C.}},
\bauthor{\bsnm{{Kumar}}, \binits{P.}},
\bauthor{\bsnm{{Chae}}, \binits{J.}},
\bauthor{\bsnm{{Liu}}, \binits{R.}},
\bauthor{\bsnm{{Wang}}, \binits{H.}}:
\byear{2012},
\batitle{{The Occurrence and Speed of CMEs Related to Two Characteristic
  Evolution Patterns of Helicity Injection in Their Solar Source Regions}}.
\bjtitle{\apj}
\bvolume{750},
\bfpage{48}.
\doiurl{10.1088/0004-637X/750/1/48}.
\adsurl{2012ApJ...750...48P}.
\end{barticle}
\endbibitem

\bibitem[\protect\citeauthoryear{{Plotnikov}
  \textit{et~al.}}{2016}]{plotnikov16}
\begin{barticle}
\bauthor{\bsnm{{Plotnikov}}, \binits{I.}},
\bauthor{\bsnm{{Rouillard}}, \binits{A.P.}},
\bauthor{\bsnm{{Davies}}, \binits{J.A.}},
\bauthor{\bsnm{{Bothmer}}, \binits{V.}},
\bauthor{\bsnm{{Eastwood}}, \binits{J.P.}},
\bauthor{\bsnm{{Gallagher}}, \binits{P.}},
\bauthor{\bsnm{{Harrison}}, \binits{R.A.}},
\bauthor{\bsnm{{Kilpua}}, \binits{E.}},
\bauthor{\bsnm{{M{\"o}stl}}, \binits{C.}},
\bauthor{\bsnm{{Perry}}, \binits{C.H.}},
\bauthor{\bsnm{{Rodriguez}}, \binits{L.}},
\bauthor{\bsnm{{Lavraud}}, \binits{B.}},
\bauthor{\bsnm{{G{\'e}not}}, \binits{V.}},
\bauthor{\bsnm{{Pinto}}, \binits{R.F.}},
\bauthor{\bsnm{{Sanchez-Diaz}}, \binits{E.}}:
\byear{2016},
\batitle{{Long-Term Tracking of Corotating Density Structures Using
  Heliospheric Imaging}}.
\bjtitle{\solphys}
\bvolume{291},
\bfpage{1853}.
\doiurl{10.1007/s11207-016-0935-9}.
\adsurl{2016SoPh..291.1853P}.
\end{barticle}
\endbibitem

\bibitem[\protect\citeauthoryear{Pontin, Priest, and
  Galsgaard}{2013}]{Pontin2013}
\begin{barticle}
\bauthor{\bsnm{Pontin}, \binits{D.I.}},
\bauthor{\bsnm{Priest}, \binits{E.R.}},
\bauthor{\bsnm{Galsgaard}, \binits{K.}}:
\byear{2013},
\batitle{On the nature of reconnection at a solar coronal null point above a
  separatrix dome}.
\bjtitle{\apj}
\bvolume{774}(\bissue{2}),
\bfpage{154}.
\burl{http://stacks.iop.org/0004-637X/774/i=2/a=154}.
\end{barticle}
\endbibitem

\bibitem[\protect\citeauthoryear{{Robbrecht} and
  {Berghmans}}{2004}]{robbrecht04}
\begin{barticle}
\bauthor{\bsnm{{Robbrecht}}, \binits{E.}},
\bauthor{\bsnm{{Berghmans}}, \binits{D.}}:
\byear{2004},
\batitle{{Automated recognition of coronal mass ejections (CMEs) in
  near-real-time data}}.
\bjtitle{\aap}
\bvolume{425},
\bfpage{1097}.
\doiurl{10.1051/0004-6361:20041302}.
\adsurl{2004A\%26A...425.1097R}.
\end{barticle}
\endbibitem

\bibitem[\protect\citeauthoryear{{Rouillard}
  \textit{et~al.}}{2017}]{rouillard17}
\begin{botherref}
\oauthor{\bsnm{{Rouillard}}, \binits{A.P.}},
\oauthor{\bsnm{{Lavraud}}, \binits{B.}},
\oauthor{\bsnm{{Genot}}, \binits{V.}},
\oauthor{\bsnm{{Bouchemit}}, \binits{M.}},
\oauthor{\bsnm{{Dufourg}}, \binits{N.}},
\oauthor{\bsnm{{Plotnikov}}, \binits{I.}},
\oauthor{\bsnm{{Pinto}}, \binits{R.F.}},
\oauthor{\bsnm{{Sanchez-Diaz}}, \binits{E.}},
\oauthor{\bsnm{{Lavarra}}, \binits{M.}},
\oauthor{\bsnm{{Penou}}, \binits{M.}},
\oauthor{\bsnm{{Jacquey}}, \binits{C.}},
\oauthor{\bsnm{{Andre}}, \binits{N.}},
\oauthor{\bsnm{{Caussarieu}}, \binits{S.}},
\oauthor{\bsnm{{Toniutti}}, \binits{J.-P.}},
\oauthor{\bsnm{{Popescu}}, \binits{D.}},
\oauthor{\bsnm{{Buchlin}}, \binits{E.}},
\oauthor{\bsnm{{Caminade}}, \binits{S.}},
\oauthor{\bsnm{{Alingery}}, \binits{P.}},
\oauthor{\bsnm{{Davies}}, \binits{J.A.}},
\oauthor{\bsnm{{Odstrcil}}, \binits{D.}},
\oauthor{\bsnm{{Mays}}, \binits{L.}}:
2017,
{A propagation tool to connect remote-sensing observations with in-situ
  measurements of heliospheric structures}.
\textit{ArXiv e-prints}.
\adsurl{2017arXiv170200399R}.
\end{botherref}
\endbibitem

\bibitem[\protect\citeauthoryear{{Sammis}, {Tang}, and
  {Zirin}}{2000}]{sammis00}
\begin{barticle}
\bauthor{\bsnm{{Sammis}}, \binits{I.}},
\bauthor{\bsnm{{Tang}}, \binits{F.}},
\bauthor{\bsnm{{Zirin}}, \binits{H.}}:
\byear{2000},
\batitle{{The Dependence of Large Flare Occurrence on the Magnetic Structure of
  Sunspots}}.
\bjtitle{\apj}
\bvolume{540},
\bfpage{583}.
\doiurl{10.1086/309303}.
\adsurl{2000ApJ...540..583S}.
\end{barticle}
\endbibitem

\bibitem[\protect\citeauthoryear{{Scherrer} \textit{et~al.}}{1995}]{scherrer95}
\begin{barticle}
\bauthor{\bsnm{{Scherrer}}, \binits{P.H.}},
\bauthor{\bsnm{{Bogart}}, \binits{R.S.}},
\bauthor{\bsnm{{Bush}}, \binits{R.I.}},
\bauthor{\bsnm{{Hoeksema}}, \binits{J.T.}},
\bauthor{\bsnm{{Kosovichev}}, \binits{A.G.}},
\bauthor{\bsnm{{Schou}}, \binits{J.}},
\bauthor{\bsnm{{Rosenberg}}, \binits{W.}},
\bauthor{\bsnm{{Springer}}, \binits{L.}},
\bauthor{\bsnm{{Tarbell}}, \binits{T.D.}},
\bauthor{\bsnm{{Title}}, \binits{A.}},
\bauthor{\bsnm{{Wolfson}}, \binits{C.J.}},
\bauthor{\bsnm{{Zayer}}, \binits{I.}},
\bauthor{\bsnm{{MDI Engineering Team}}}:
\byear{1995},
\batitle{{The Solar Oscillations Investigation - Michelson Doppler Imager}}.
\bjtitle{\solphys}
\bvolume{162},
\bfpage{129}.
\doiurl{10.1007/BF00733429}.
\adsurl{1995SoPh..162..129S}.
\end{barticle}
\endbibitem

\bibitem[\protect\citeauthoryear{{Scherrer} \textit{et~al.}}{2012}]{scherrer12}
\begin{barticle}
\bauthor{\bsnm{{Scherrer}}, \binits{P.H.}},
\bauthor{\bsnm{{Schou}}, \binits{J.}},
\bauthor{\bsnm{{Bush}}, \binits{R.I.}},
\bauthor{\bsnm{{Kosovichev}}, \binits{A.G.}},
\bauthor{\bsnm{{Bogart}}, \binits{R.S.}},
\bauthor{\bsnm{{Hoeksema}}, \binits{J.T.}},
\bauthor{\bsnm{{Liu}}, \binits{Y.}},
\bauthor{\bsnm{{Duvall}}, \binits{T.L.}},
\bauthor{\bsnm{{Zhao}}, \binits{J.}},
\bauthor{\bsnm{{Title}}, \binits{A.M.}},
\bauthor{\bsnm{{Schrijver}}, \binits{C.J.}},
\bauthor{\bsnm{{Tarbell}}, \binits{T.D.}},
\bauthor{\bsnm{{Tomczyk}}, \binits{S.}}:
\byear{2012},
\batitle{{The Helioseismic and Magnetic Imager (HMI) Investigation for the
  Solar Dynamics Observatory (SDO)}}.
\bjtitle{\solphys}
\bvolume{275},
\bfpage{207}.
\doiurl{10.1007/s11207-011-9834-2}.
\adsurl{2012SoPh..275..207S}.
\end{barticle}
\endbibitem

\bibitem[\protect\citeauthoryear{Scherrer \textit{et~al.}}{2012}]{sdo_hmi}
\begin{barticle}
\bauthor{\bsnm{Scherrer}, \binits{P.H.}},
\bauthor{\bsnm{Schou}, \binits{J.}},
\bauthor{\bsnm{Bush}, \binits{R.I.}},
\bauthor{\bsnm{Kosovichev}, \binits{A.G.}},
\bauthor{\bsnm{Bogart}, \binits{R.S.}},
\bauthor{\bsnm{Hoeksema}, \binits{J.T.}}:
\byear{2012},
\batitle{The helioseismic and magnetic imager (hmi) investigation for the solar
  dynamics observatory (sdo)}.
\bjtitle{\solphys}
\bvolume{275},
\bfpage{207}.
\doiurl{10.1007/s11207-011-9834-2}.
\burl{http://dx.doi.org/10.1007/s11207-011-9834-2}.
\end{barticle}
\endbibitem

\bibitem[\protect\citeauthoryear{{Schrijver}}{2007a}]{schrijver07}
\begin{barticle}
\bauthor{\bsnm{{Schrijver}}, \binits{C.J.}}:
\byear{2007}a,
\batitle{{A Characteristic Magnetic Field Pattern Associated with All Major
  Solar Flares and Its Use in Flare Forecasting}}.
\bjtitle{\apjl}
\bvolume{655},
\bfpage{L117}.
\doiurl{10.1086/511857}.
\adsurl{2007ApJ...655L.117S}.
\end{barticle}
\endbibitem

\bibitem[\protect\citeauthoryear{{Schrijver}}{2007b}]{2007ApJ...655L.117S}
\begin{barticle}
\bauthor{\bsnm{{Schrijver}}, \binits{C.J.}}:
\byear{2007}b,
\batitle{{A Characteristic Magnetic Field Pattern Associated with All Major
  Solar Flares and Its Use in Flare Forecasting}}.
\bjtitle{\apjl}
\bvolume{655},
\bfpage{L117}.
\doiurl{10.1086/511857}.
\adsurl{2007ApJ...655L.117S}.
\end{barticle}
\endbibitem

\bibitem[\protect\citeauthoryear{{Shibata} and
  {Magara}}{2011}]{2011LRSP....8....6S}
\begin{barticle}
\bauthor{\bsnm{{Shibata}}, \binits{K.}},
\bauthor{\bsnm{{Magara}}, \binits{T.}}:
\byear{2011},
\batitle{{Solar Flares: Magnetohydrodynamic Processes}}.
\bjtitle{Living Rev. Solar Phys.}
\bvolume{8},
\bfpage{6}.
\adsurl{http://cdsads.u-strasbg.fr/abs/2011LRSP....8....6S}.
\end{barticle}
\endbibitem

\bibitem[\protect\citeauthoryear{{Singh} and
  {Badruddin}}{2006}]{2006JASTP..68..803S}
\begin{barticle}
\bauthor{\bsnm{{Singh}}, \binits{Y.P.}},
\bauthor{\bsnm{{Badruddin}}}:
\byear{2006},
\batitle{{Statistical considerations in superposed epoch analysis and its
  applications in space research}}.
\bjtitle{\jastp}
\bvolume{68},
\bfpage{803}.
\doiurl{10.1016/j.jastp.2006.01.007}.
\adsurl{2006JASTP..68..803S}.
\end{barticle}
\endbibitem

\bibitem[\protect\citeauthoryear{{Tiwari} \textit{et~al.}}{2015}]{tiwari15}
\begin{barticle}
\bauthor{\bsnm{{Tiwari}}, \binits{S.K.}},
\bauthor{\bsnm{{Falconer}}, \binits{D.A.}},
\bauthor{\bsnm{{Moore}}, \binits{R.L.}},
\bauthor{\bsnm{{Venkatakrishnan}}, \binits{P.}},
\bauthor{\bsnm{{Winebarger}}, \binits{A.R.}},
\bauthor{\bsnm{{Khazanov}}, \binits{I.G.}}:
\byear{2015},
\batitle{{Near-Sun speed of CMEs and the magnetic nonpotentiality of their
  source active regions}}.
\bjtitle{\grl}
\bvolume{42},
\bfpage{5702}.
\doiurl{10.1002/2015GL064865}.
\adsurl{2015GeoRL..42.5702T}.
\end{barticle}
\endbibitem

\bibitem[\protect\citeauthoryear{{Venkatakrishnan} and
  {Ravindra}}{2003}]{ventakrishnan03}
\begin{barticle}
\bauthor{\bsnm{{Venkatakrishnan}}, \binits{P.}},
\bauthor{\bsnm{{Ravindra}}, \binits{B.}}:
\byear{2003},
\batitle{{Relationship between CME velocity and active region magnetic
  energy}}.
\bjtitle{\grl}
\bvolume{30},
\bfpage{2181}.
\doiurl{10.1029/2003GL018100}.
\adsurl{2003GeoRL..30.2181V}.
\end{barticle}
\endbibitem

\bibitem[\protect\citeauthoryear{{Vourlidas} and {Howard}}{2006}]{vourlidas06}
\begin{barticle}
\bauthor{\bsnm{{Vourlidas}}, \binits{A.}},
\bauthor{\bsnm{{Howard}}, \binits{R.A.}}:
\byear{2006},
\batitle{{The Proper Treatment of Coronal Mass Ejection Brightness: A New
  Methodology and Implications for Observations}}.
\bjtitle{\apj}
\bvolume{642},
\bfpage{1216}.
\doiurl{10.1086/501122}.
\adsurl{2006ApJ...642.1216V}.
\end{barticle}
\endbibitem

\bibitem[\protect\citeauthoryear{Wang \textit{et~al.}}{2014}]{Wang2014}
\begin{barticle}
\bauthor{\bsnm{Wang}, \binits{S.}},
\bauthor{\bsnm{Liu}, \binits{C.}},
\bauthor{\bsnm{Deng}, \binits{N.}},
\bauthor{\bsnm{Wang}, \binits{H.}}:
\byear{2014},
\batitle{Sudden photospheric motion and sunspot rotation associated with the
  x2.2 flare on 2011 february 15}.
\bjtitle{\apjl}
\bvolume{782}(\bissue{2}),
\bfpage{L31}.
\burl{http://stacks.iop.org/2041-8205/782/i=2/a=L31}.
\end{barticle}
\endbibitem

\bibitem[\protect\citeauthoryear{{Wang} \textit{et~al.}}{2011}]{wang11}
\begin{barticle}
\bauthor{\bsnm{{Wang}}, \binits{Y.}},
\bauthor{\bsnm{{Chen}}, \binits{C.}},
\bauthor{\bsnm{{Gui}}, \binits{B.}},
\bauthor{\bsnm{{Shen}}, \binits{C.}},
\bauthor{\bsnm{{Ye}}, \binits{P.}},
\bauthor{\bsnm{{Wang}}, \binits{S.}}:
\byear{2011},
\batitle{{Statistical study of coronal mass ejection source locations:
  Understanding CMEs viewed in coronagraphs}}.
\bjtitle{\jgr (Space Physics)}
\bvolume{116},
\bfpage{A04104}.
\doiurl{10.1029/2010JA016101}.
\adsurl{2011JGRA..116.4104W}.
\end{barticle}
\endbibitem

\bibitem[\protect\citeauthoryear{{Webb} and {Howard}}{2012}]{webb12}
\begin{barticle}
\bauthor{\bsnm{{Webb}}, \binits{D.F.}},
\bauthor{\bsnm{{Howard}}, \binits{T.A.}}:
\byear{2012},
\batitle{{Coronal Mass Ejections: Observations}}.
\bjtitle{Living Reviews in Solar Physics}
\bvolume{9},
\bfpage{3}.
\doiurl{10.12942/lrsp-2012-3}.
\adsurl{2012LRSP....9....3W}.
\end{barticle}
\endbibitem

\bibitem[\protect\citeauthoryear{Yang \textit{et~al.}}{2004}]{Yang2004}
\begin{barticle}
\bauthor{\bsnm{Yang}, \binits{G.}},
\bauthor{\bsnm{Xu}, \binits{Y.}},
\bauthor{\bsnm{Cao}, \binits{W.}},
\bauthor{\bsnm{Wang}, \binits{H.}},
\bauthor{\bsnm{Denker}, \binits{C.}},
\bauthor{\bsnm{Rimmele}, \binits{T.R.}}:
\byear{2004},
\batitle{Photospheric shear flows along the magnetic neutral line of active
  region 10486 prior to an x10 flare}.
\bjtitle{\apjl}
\bvolume{617}(\bissue{2}),
\bfpage{L151}.
\burl{http://stacks.iop.org/1538-4357/617/i=2/a=L151}.
\end{barticle}
\endbibitem

\bibitem[\protect\citeauthoryear{{Yashiro} \textit{et~al.}}{2005}]{yashiro05}
\begin{barticle}
\bauthor{\bsnm{{Yashiro}}, \binits{S.}},
\bauthor{\bsnm{{Gopalswamy}}, \binits{N.}},
\bauthor{\bsnm{{Akiyama}}, \binits{S.}},
\bauthor{\bsnm{{Michalek}}, \binits{G.}},
\bauthor{\bsnm{{Howard}}, \binits{R.A.}}:
\byear{2005},
\batitle{{Visibility of coronal mass ejections as a function of flare location
  and intensity}}.
\bjtitle{\jgr (Space Physics)}
\bvolume{110},
\bfpage{A12S05}.
\doiurl{10.1029/2005JA011151}.
\adsurl{2005JGRA..11012S05Y}.
\end{barticle}
\endbibitem

\bibitem[\protect\citeauthoryear{{Yashiro} \textit{et~al.}}{2008}]{yashiro08}
\begin{barticle}
\bauthor{\bsnm{{Yashiro}}, \binits{S.}},
\bauthor{\bsnm{{Michalek}}, \binits{G.}},
\bauthor{\bsnm{{Akiyama}}, \binits{S.}},
\bauthor{\bsnm{{Gopalswamy}}, \binits{N.}},
\bauthor{\bsnm{{Howard}}, \binits{R.A.}}:
\byear{2008},
\batitle{{Spatial Relationship between Solar Flares and Coronal Mass
  Ejections}}.
\bjtitle{\apj}
\bvolume{673},
\bfpage{1174}.
\doiurl{10.1086/524927}.
\adsurl{2008ApJ...673.1174Y}.
\end{barticle}
\endbibitem

\bibitem[\protect\citeauthoryear{{Youssef}}{2012}]{youssef12}
\begin{barticle}
\bauthor{\bsnm{{Youssef}}, \binits{M.}}:
\byear{2012},
\batitle{{On the relation between the CMEs and the solar flares}}.
\bjtitle{NRIAG Journal of Astronomy and Geophysics}
\bvolume{1},
\bfpage{172}.
\doiurl{10.1016/j.nrjag.2012.12.014}.
\adsurl{2012JAsGe...1..172Y}.
\end{barticle}
\endbibitem

\bibitem[\protect\citeauthoryear{{Yurchyshyn}
  \textit{et~al.}}{2005}]{yurchyshyn05}
\begin{barticle}
\bauthor{\bsnm{{Yurchyshyn}}, \binits{V.}},
\bauthor{\bsnm{{Yashiro}}, \binits{S.}},
\bauthor{\bsnm{{Abramenko}}, \binits{V.}},
\bauthor{\bsnm{{Wang}}, \binits{H.}},
\bauthor{\bsnm{{Gopalswamy}}, \binits{N.}}:
\byear{2005},
\batitle{{Statistical Distributions of Speeds of Coronal Mass Ejections}}.
\bjtitle{\apj}
\bvolume{619},
\bfpage{599}.
\doiurl{10.1086/426129}.
\adsurl{2005ApJ...619..599Y}.
\end{barticle}
\endbibitem

\bibitem[\protect\citeauthoryear{Zheng}{2013}]{zheng13}
\begin{barticle}
\bauthor{\bsnm{Zheng}, \binits{Y.}}:
\byear{2013},
\batitle{Improving cme forecasting capability: An urgent need}.
\bjtitle{Space Weather}
\bvolume{11}(\bissue{11}),
\bfpage{641}.
\bcomment{2013SW001004}.
\doiurl{10.1002/2013SW001004}.
\burl{http://dx.doi.org/10.1002/2013SW001004}.
\end{barticle}
\endbibitem

\bibitem[\protect\citeauthoryear{Zuccarello
  \textit{et~al.}}{2014}]{Zuccarello2014}
\begin{barticle}
\bauthor{\bsnm{Zuccarello}, \binits{F.P.}},
\bauthor{\bsnm{Seaton}, \binits{D.B.}},
\bauthor{\bsnm{Mierla}, \binits{M.}},
\bauthor{\bsnm{Poedts}, \binits{S.}},
\bauthor{\bsnm{Rachmeler}, \binits{L.A.}},
\bauthor{\bsnm{Romano}, \binits{P.}},
\bauthor{\bsnm{Zuccarello}, \binits{F.}}:
\byear{2014},
\batitle{Observational evidence of torus instability as trigger mechanism for
  coronal mass ejections: The 2011 august 4 filament eruption}.
\bjtitle{\apjl}
\bvolume{785}(\bissue{2}),
\bfpage{88}.
\burl{http://stacks.iop.org/0004-637X/785/i=2/a=88}.
\end{barticle}
\endbibitem

\end{thebibliography}
%

%
%
%

\end{article} 
\end{document}